\documentclass[12pt]{article}
\usepackage[english]{babel}    
\usepackage[ansinew]{inputenc} 
\usepackage[T1]{fontenc}       
\usepackage{lmodern,microtype} 
\usepackage{amsmath,amsthm,amssymb,amsfonts} 
\usepackage{dsfont,mathrsfs,ushort} 

\usepackage{titlesec,titling}  
\usepackage[nohead]{geometry}  
\usepackage{setspace,caption}  
\usepackage{enumitem,booktabs} 
\usepackage[comma]{natbib} 
\usepackage{tikz}
\usepackage{hyperref}
\usepackage{subcaption}
\usepackage{longtable}
\usepackage[flushleft]{threeparttable}

\setenumerate{label=\small(\roman*)}

\hypersetup{colorlinks=true,pdfnewwindow=true,pdfstartview=FitH,%
	pdftitle="UnobservedCS",pdfauthor="Aguiar Kashaev",%
	urlcolor=blue!90!red!45!black,citecolor=blue!90!red!45!black,linkcolor=red!90!black}
\DeclareCaptionFont{fancy}{\bfseries\sffamily}
\allowdisplaybreaks

\usepackage{graphicx}

\titleformat{\section}[block]{\centering\large\bfseries\sffamily}{\thesection.}{0.5em}{}
\titleformat{\subsection}[block]{\flushleft\bfseries}{\thesubsection.}{0.5em}{}
\titleformat{\subsection}[block]{\flushleft\bfseries\sffamily}{\thesubsection.}{0.5em}{}
\titleformat{\subsubsection}[runin]{\normalsize\bfseries\sffamily}{\bfseries\upshape\sffamily\thesubsubsection.}{0.5em}{}[.--\:]
\renewcommand{\thesubsubsection}{\arabic{section}.\arabic{subsection}.\arabic{subsubsection}}
\titlespacing{\section}{0ex}{10ex}{5ex}
\titlespacing{\subsection}{0in}{6ex}{3ex}
\titlespacing{\subsubsection}{0mm}{2ex}{0.5em}
\pretitle{\begin{center}\LARGE\bfseries\sffamily}
\posttitle{\par\end{center}\vskip 0.5em}
\preauthor{\begin{center} \large \lineskip 0.5em\begin{tabular}[t]{c}}
\postauthor{\end{tabular}\par\end{center}}
\predate{\begin{center}\small}
\postdate{\par\end{center}}
\providecommand{\abstitle}[1]{{\par\vspace*{2ex}\small\bfseries\sffamily #1}\hspace*{1ex}}
\renewenvironment{abstract}%
{\begin{center}\begin{minipage}{0.8\linewidth}%
			\setlength{\parindent}{0.0em}\abstitle{Abstract}\small}%
		{\end{minipage}\end{center}\vfill\clearpage}

\DeclareMathOperator*{\argmax}{arg\,max}
\DeclareMathOperator*{\argmin}{arg\,min}

\providecommand{\Char}[1]{\mathds{1}\left(\,#1\,\right)}
\providecommand{\Real}{{\mathds{R}}}

\providecommand{\tr}{^{{\sf T}}}

\providecommand{\rand}[1]{\mathbf{#1}}
\providecommand{\rands}[1]{\boldsymbol{#1}}

\providecommand{\Prob}[1]{\mathds{P}\left(#1\right)}

\providecommand{\abs}[1]{\left\lvert#1\right\rvert}

\newcommand{\x}{x}\newcommand{\X}{X}
\newcommand{\y}{y}
\newcommand{\z}{z}\newcommand{\Z}{Z}
\newcommand{\w}{w}

 \renewcommand{\u}{u}
 
\newcommand{\D}{D}

\newcommand{\rumo}{\ensuremath{\mathrm{RUM}}}
  \theoremstyle{remark}
  
  \theoremstyle{plain}
  \newtheorem{lem}{\protect\lemmaname}
  \theoremstyle{definition}
  
\theoremstyle{plain}
\newtheorem{thm}{\protect\theoremname}

  \theoremstyle{plain}
  \newtheorem{cor}{\protect\corollaryname}
 \theoremstyle{definition}
  \newtheorem{example}{\protect\examplename}
  \theoremstyle{plain}
  \newtheorem{assumption}{\protect\assumptionname}
\newtheorem{prop}{Proposition}

  \providecommand{\assumptionname}{Assumption}

  \providecommand{\definitionname}{Definition}
  \providecommand{\lemmaname}{Lemma}
  
  \providecommand{\remarkname}{Remark}
\providecommand{\corollaryname}{Corollary}
\providecommand{\theoremname}{Theorem}
\providecommand{\examplename}{Example}
\newcounter{aux}
\newcounter{egmain}
\newtheoremstyle{example}
{\topsep} {\topsep}%
{\upshape}
{}
{\bfseries\scshape}
{.}
{1em}
{}

\theoremstyle{example}

\newtheoremstyle{example_contd}
{\topsep} {\topsep}%
{\upshape}
{}
{\bfseries\scshape}
{.}
{1em}
{\thmname{#1} \thmnumber{ #2}\thmnote{#3} (continued)}

\theoremstyle{example_contd}

\makeatother

\begin{document}
\title{Identification and Estimation of Discrete Choice Models with Unobserved Choice Sets\thanks{The ``\textcircled{r}'' symbol indicates that the authors' names are in certified random order, as described by \citet{ray2018certified}. We would like to thank Roy Allen, Daniel Chaves, Mingshi Kang, Yuichi Kitamura, Mathieu Marcoux, Salvador Navarro, Joris Pinkse, David Rivers, Bruno Salcedo, Susanne Schennach, Tomasz Strzalecki, and David Wei for helpful comments and suggestions. We are grateful to the editor, the associate editor, and three referees for constructive comments and suggestions. We gratefully acknowledge financial support from the Western Social Science Faculty Grant (FRDF R5533A02) and Social Sciences and Humanities Research Council Insight Development Grant. Researchers own analyses calculated (or derived) based in part on data from Nielsen Consumer LLC and marketing databases provided through the NielsenIQ Datasets at the Kilts Center for Marketing Data Center at The University of Chicago Booth School of Business. The conclusions drawn from the NielsenIQ data are those of the researchers and do not reflect the views of Nielsen. Nielsen is not responsible for, had no role in, and was not involved in analyzing and preparing the results reported herein.}}

\author{ 
	Victor H. Aguiar \textcircled{r}
	Nail Kashaev\thanks{Aguiar: Department of Economics, University of Western Ontario; \href{mailto:vaguiar@uwo.ca}{vaguiar@uwo.ca}. Kashaev: Department of Economics, University of Western Ontario; \href{mailto:nkashaev@uwo.ca}{nkashaev@uwo.ca}.}
	}
\date{This version: May 14, 2021 / First Version: July 9, 2019}
\maketitle
\begin{abstract} 
We propose a framework for nonparametric identification and estimation of discrete choice models with unobserved choice sets. We recover the joint distribution of choice sets and preferences from a panel dataset on choices. We assume that either the latent choice sets are \textit{sparse} or that the panel is sufficiently long. Sparsity requires the number of possible choice sets to be relatively small. It is satisfied, for instance, when the choice sets are nested, or when they form a partition.
Our estimation procedure is computationally fast and uses mixed-integer optimization to recover the sparse support of choice sets. Analyzing the ready-to-eat cereal industry using a household scanner dataset, we find that ignoring the unobservability of choice sets can lead to biased estimates of preferences due to significant latent heterogeneity in choice sets. 
\end{abstract}

\section{Introduction}\label{section:Introduction}
This paper considers the nonparametric identification and estimation of discrete choice models when the choice set that decision makers (DMs) face are unobserved by the researcher.\footnote{See \citet{manski1977structure} for the early treatment of the problem.} We show how to nonparametrically identify and estimate the joint distribution of latent choice sets and preferences when we observe a panel dataset on choices. We allow random preferences to have different distributions across time and to be correlated with choice sets after conditioning on observed covariates. We apply our methodology to analyze the ready-to-eat (RTE) cereal industry in the USA using a household scanner dataset (Nielsen Homescan). Our empirical findings suggest there is substantial latent choice set heterogeneity that can bias estimates of parameters of interest if not taken into account.
\par
The classical nonparametric treatment of discrete choice under random utility \citep{mcfadden1990stochastic}\footnote{From a decision-theoretic tradition random utility was initially studied by \citet{block_random_1960} and \citet{falmagne_representation_1978}.} uses exogenous choice set variation to identify the distribution of preferences nonparametrically. However, the researcher usually does not observe the choice sets from which DMs pick their most preferred alternative. As a response to this lack of observability, researchers usually impose parametric restrictions on the distribution of preferences and the distribution of choice sets, or assume that every DM faces the same choice set \citep{hickman201613}. These assumptions are problematic as they may lead to inconsistent estimation of preferences. 
\par
We overcome the above issue by exploiting the panel structure of our data. That is, we use variation in choices of the same DM in different time periods. Assuming that the choice set of the same DM remains the same over time, the observed sequence of choices from the same choice set reveals information about it. Intuitively, if we observe a DM  who chooses repeatedly either alternative $1$ or alternative $2$, then we can conclude that, most likely, this DM considers only these two alternatives. Another DM who chooses alternatives $1$, $2$, and $3$, most likely, has a bigger consideration set than the first DM. This variation in choices across time allows us to pin down the choice sets. After the choice sets are identified, the variation in choices within a given choice set allows us to identify the distribution of choices conditional on choice sets.\footnote{Our work provides a methodological bridge between the decision-theoretic literature on stochastic choice that has been based fundamentally on choice set variation (e.g., \citet{luce1959individual, block_random_1960, falmagne_representation_1978, mcfadden1990stochastic, gul2006random, manzini2014stochastic, fudenberg2015stochastic}, and \citet{brady2016menudependent}) and the discrete choice literature that has exploited covariate variation to identify the parametric distribution of preferences \citep{train2009discrete}.}  
\par
Formally, to establish our identification result, we show that the problem of discrete choice with unobserved choice sets can be framed as a finite mixture, thus permitting us to use recent advances in identification of these models.\footnote{See, for instance, \citet{hall2003nonparametric,hu2008identification,kasahara2009nonparametric, bonhomme2016non, kitamura2018finitemix}, and references therein.} 
In particular, we base our identification strategy on the insights from discrete nonclassical measurement error results in \citet{hu2008identification}.\footnote{For recent applications in the context of auctions and discrete games see \citet{hu2013identification, xiao2018identification}, and \citet{luo2018unobserved}.} We require at least three observed choices from the same DM over time that are conditionally independent conditional on the unobserved choice set and observed covariates. Using these three choices, we show that under the standard linear independence (rank) condition the identification of both the distribution of choice sets and the distribution of preferences is possible. We differ from \citet{hu2008identification} in that we do not need to impose any strict monotonicity condition, and we do not need to know the number of possible choice sets since we can identify and estimate it. Strict monotonicity restrictions are usually imposed to match anonymous functions to latent states. In our setting, the latent choice sets have a structural interpretation and we do not need to rank them. 
\par
The structural interpretation of the latent choice sets also allows us to establish two new sufficient conditions for the linear independence condition. The first condition imposes no restrictions on the distribution of preferences or choice sets but requires the panel to be sufficiently long. The second condition requires less data but imposes \emph{sparsity} on the support of the latent choice sets. Given that the number of all possible choice sets grows exponentially with the number of alternatives, the sparsity condition allows for substantial dimensionality reduction.
\par
We also provide a new consistent computationally efficient nonparametric estimator of the distribution of choice sets and choices conditional on choice sets. Our estimator is a two-step estimator. On the first step, we consistently estimate choice sets using our identification result. However, this estimator may not perform very well in finite samples. That is why, in the second step, we regularize it to achieve better finite sample properties. Using sparsity, we show that the problem of estimating a small number of sets can be cast to \emph{the best subset regression problem} (see, for instance, \citealp{bertsimas2016best}). As a result, the problem turns out to be a mixed-integer program that can be solved very quickly with modern optimization routines.
\par
We apply our estimator to the Nielsen Homescan dataset. We study the RTE cereal market. This market has been studied in \citet{nevo2001measuring} using aggregate datasets. The RTE cereal market is known to be highly concentrated, with high differentiation, large advertisement-to-sale ratios, and product innovation. All of these factors suggest high variability of choice sets because of consumer loyalty, geographical variation in product availability, and targeted advertisement campaigns.  We exploit the high frequency of the panel in the Nielsen Homescan with roughly weekly time variation to uncover substantial heterogeneity in choice set variation across markets. Furthermore, we find evidence in favor of our sparsity assumption and show that ignoring this latent choice set heterogeneity leads to biased estimates of price elasticities in a simple model of demand in the spirit of \citet{berry1995automobile} and \citet{nevo2001measuring}.     
\par
The closest work to ours is \citet{crawford2019demand}. We differ from their work in several respects. \citet{crawford2019demand} mainly consider settings where either choice sets do not change (stable choice sets) or become larger over time (growing choice sets). They do not put restrictions on how choice sets for different DMs are related. We work with settings where choice sets DMs face are stable across time, but sparse across DMs. Next, our approach is fully nonparametric while \citet{crawford2019demand} work with the stylized multinomial logit model of choice and impose parametric restrictions on the choice set distribution.\footnote{See also  \citet{goeree2008limited} and \citet{barseghyan2019discrete} for applications of consideration sets driven by item-dependent attention.} 
We also differ in that we allow for the endogenous selection of DMs into choice sets. That is, we allow for correlation between preferences and choice sets even after conditioning on covariates. This is an important distinction in our empirical application, since we find that consumers with different choice sets exhibit different choice behavior.
\citet{fieldMM19} recover jointly the distribution of preferences and the distribution of consideration sets under parametric restrictions on the distribution of consideration sets. We do not require such parametric restrictions.\footnote{In an alternative strand of the literature, \citet{abaluck2017consumers} exploit parametric restrictions on preferences and consideration to achieve identification without panel datasets and exclusion restrictions by assuming asymmetries in the substitution matrix. \citet{allen2019revealed} show how the important models of limited consideration of \citet{manzini2014stochastic} and \citet{brady2016menudependent} can fail to satisfy this property.}
\citet{lu2014moment} and \citet{molinari2018} use only cross-sectional variation and set-identify the parametric distribution of preferences only.\footnote{\citet{lu2014moment} also provides a set of conditions that ensure that a system of moment inequalities he builds uniquely identifies the parameter of interest.} We point-identify and estimate the joint distribution of preferences and choice sets nonparametrically using panel data.
\par
Section~\ref{sec:model} presents the model. Section~\ref{sec: identification} contains our main identification result. In Section~\ref{sec: estimation} we describe our estimation procedure. Section~\ref{sec:simulations} assesses the performance of our estimator in simulations. In Section~\ref{sec: empirical application} we present our empirical application. Finally, Section~\ref{sec:conclusion} concludes. All proofs can be found in Appendix~\ref{app:proofs}. Appendices~\ref{app: MC simulations} and~\ref{app: empirical application} contain additional simulations and estimation results.

\section{Model}\label{sec:model}
\subsection{Choice Sets and Preferences}
We consider an environment where at time $t=0$ a DM is faced (at random) with a finite choice set $\rand{D}$ and then at every $t\in\mathcal{T}=\{1,2,\dots,T\}$ chooses the alternative $y_t$ from $\rand{D}$ that maximizes her random preferences.\footnote{We use boldface font (e.g. $\rand{D}$) to denote random objects and regular font (e.g. $D$) for deterministic ones.} The preferences at every moment of time are captured by the random strict preference orders represented by random utility functions $\rand{u}=\{\rand{u}_t\}_{t\in\mathcal{T}}$ that are defined over some grand choice set that contains $\rand{D}$ with probability $1$. Without loss of generality, we assume that the grand choice set is $\mathcal{Y}=\{1,2,\dots,Y\}$, where $Y$ is a finite constant.
\par
Let $\rand{x}\in X\subseteq\Real^{d_x}$ denote the vector of observed covariates. The set of covariates depends on a particular application and can include DM-specific characteristics (e.g., age and gender) and choice-problem-specific characteristics (e.g., zip code, location of the store, day of the year, month, or time of the day).

\begin{assumption}[Observables]\label{ass:observables}
The researcher observes (can consistently estimate) the joint distribution of $(\rand{y}_t)_{t\in\mathcal{T}}$ and $\rand{x}$.
\end{assumption}

\setcounter{egmain}{\value{example}}
\begin{example}\label{eg:main}
Suppose that $Y$ brands of a product (e.g. cereal) are available in a given location (market). Let $\rand{x}=((\tilde{\rand{x}}\tr_{y,t})\tr_{y\in\mathcal{Y},t\in\mathcal{T}},(\rand{r}_{t})_{t\in\mathcal{T}})\tr$ be the vector of observed covariates, where $\tilde{\rand{x}}_{y,t}$ is the vector of characteristics of product $y$ at time $t$ (e.g., price and package size); $\rand{r}_{t}$ is the vector of characteristics of a DM (e.g., age and income level) and market (e.g., market identifier). The DM draws a latent choice set $\rand{D}\subseteq\mathcal{Y}$ and purchases a product $\rand{y}_t$ from that set at every time period $t$. The analyst observes a sample of $n$ independently and identically distributed (i.i.d.) across DMs observations $\left\{\left(\rand{y}^{(i)}_{t}\right)_{t\in\mathcal{T}},\rand{x}^{(i)}\right\}_{i=1}^{n}$ drawn from a joint distribution of $(\rand{y}_{t})_{t\in\mathcal{T}}$ and $\rand{x}$ (a panel dataset).
\end{example}

While we work with the environments where the choice sets do not change over time (the random choice set $\rand{D}$ is not indexed by $t$), we still allow agents to make different choices in different time periods. Following the classical treatment of the Random Utility Model (RUM, \citealp{mcfadden1973conditional}) with \emph{observed} choice sets we make the following assumption. 

\begin{assumption}[Conditionally Independent Preferences]\label{ass: CIP}
Conditional on the realization of covariates and the choice set, preferences are independent across time. That is,
\[
\rand{u}_t\perp\rand{u}_s\mid (\rand{x}=x,\rand{D}=D),
\]
for all $t,s\in \mathcal{T}$, $\x\in X$, and $D\in\mathcal{D}_\x$, where $\mathcal{D}_\x$ is the conditional support of $\rand{D}$ conditional on $\rand{\x}=\x$.
\end{assumption}

We emphasize that Assumption~\ref{ass: CIP} does not restrict the dependence structure between $\rand{u}$, $\rand{x}$, and $\rand{D}$. In particular, we allow preferences to be correlated with choice sets. (See examples below and our empirical application). Moreover, preferences $\{\rand{u}_t\}_{t\in\mathcal{T}}$ and, thus, choices are correlated across time through the latent choice set $\rand{D}$ and covariates $\rand{x}$. Assumption~\ref{ass: CIP} is standard in the analysis of differentiated products demand systems using market and individual level data (e.g., \citealp{berry1995automobile, nevo2000practitioner, nevo2001measuring, lu2014moment, crawford2019demand}).\footnote{In the literature that uses market-level data, the independent markets are often defined using a time interval (e.g., week, quarter, or year), location (e.g, town or zip-code), or both.}  In Section~\ref{sec: markov}, we relax this conditional independence assumption to allow some dynamics due to unobserved heterogeneity on top of $\rand{D}$.

\setcounter{aux}{\value{example}}
\setcounter{example}{\value{egmain}}
\begin{example}(continued)
	Suppose that a DM, who faces a choice set $\rand{D}=D$, obtains the following utility from purchasing product $y$ at time $t$:
	\[
	\rand{u}_{y,D,t}=\alpha_{y,D,t}(\rand{x})+\rands{\varepsilon}_{y,D,t},
	\]
	where $\{\alpha_{y,D,t}\}_{y\in\mathcal{Y},D\in\mathcal{D}_x,t\in\mathcal{T}}$ are unknown functions that map $X$ to $\Real$; and $\{\rands{\varepsilon}_{y,D,t}\}_{y\in\mathcal{Y},D\in\mathcal{D}_x,t\in\mathcal{T}}$ are taste shocks that are independent across $t$, but potentially correlated across $y$ and $D$. Moreover, $\rand{\varepsilon}$s are allowed to be correlated with $\rand{x}$. Functions $\alpha_{y,D,t}$ may include unknown product-time-choice set fixed effects. In this example, Assumption~\ref{ass: CIP} is satisfied. Also note that two DMs may get different utilities from the same product at the same moment of time if their choice sets are different.
\end{example}
\setcounter{example}{\value{aux}}

Assumptions~\ref{ass: CIP} together with the assumption that DMs are maximizing utility implies that, after the choice set is realized, the choices of DMs are consistent with the classic RUM. In other words, after conditioning on the choice set and covariates, choices of the DM are independent, but not necessarily identically distributed. Indeed, we can define
\[
y^{\rumo}(D,\u_t)=\argmax_{y'\in D}u_t(y')
\]
for every choice set $D$ and utility function $u_t$, and
\[
F_t^{\rumo}(y\mid D,x)=\Prob{\y=y^{\rumo}(\rand{D},\rand{u}_t)\mid \rand{\x}=\x,\rand{D}=D}
\]
for every choice $y$, covariate $x$, and choice set $D\in \mathcal{D}_x$. $F_t^{\rumo}$ is a conditional probability mass function that captures the random utility maximization part of DMs choices. Thus, in terms of random choices, Assumption~\ref{ass: CIP} implies that for all $t$ and $s$, $t\neq s$, the observed choices have to satisfy
\begin{equation}\label{eq: CIC}
\begin{aligned}
&\rand{y}_t\mid (\rand{\D}=\D, \rand{x}=x) \sim \:F_t^{\rumo}(\cdot\mid \D,\x),\\
&\rand{y}_t\perp\rand{y}_s\mid (\rand{\D}=\D, \rand{x}=x).
\end{aligned}
\end{equation}
As a result, we can rewrite the conditional distribution of observed choices at any $t\in\mathcal{T}$ as a finite mixture model:
\begin{align*}
\Pr(\rand{\y}_t=\y\mid \rand{\x}=x)=\sum_{\D\in\mathcal{\D}_x}m(\D\mid \x)F_t^{\rumo}(y\mid \D,\x)
\end{align*}
for all $x$ and $y$. Using the data on choices and covariates, the researcher is interested in recovering the conditional distribution of choice sets captured by $m$ and the random utility maximization aspects of the model captured by $F^\rumo=\{F_t^{\rumo}\}_{t\in\mathcal{T}}$.
\par
Next we impose the following regularity condition on $F_t^{\rumo}$.

\begin{assumption}[Full Support]\label{ass:regularity}
For every $\x\in\X$, $D\in\mathcal{D}_x$, $t\in\mathcal{T}$, and $y\in D$, 
\[
F_t^{\rumo}(\y\mid  D,\x)>0.
\]
\end{assumption}
Assumption~\ref{ass:regularity} is a standard assumption in discrete choice literature: every alternative in every choice set is chosen with positive probability. \citet{mcfadden1973conditional} pointed out that in finite samples, Assumption~\ref{ass:regularity} is not testable, since zero market shares are not distinguishable from arbitrarily small but positive market shares. Additionally, if an alternative is never observed in the data, then it may be that this alternative either does not belong to any choice set or is always dominated by another alternative. Assumption~\ref{ass:regularity} excludes such cases.

\setcounter{aux}{\value{example}}
\setcounter{example}{\value{egmain}}
\begin{example}(continued)
	If the support of random vector $(\rands{\varepsilon}_{D,t,y})_{y\in\mathcal{Y}}$ is $\Real^{Y}$ for all $D,t$, then Assumption~\ref{ass:regularity} is satisfied.
\end{example}
\setcounter{example}{\value{aux}}

\section{Identification}\label{sec: identification}
In this section, we provide our main identification result and extend our model to allow choices of the same DM to be correlated across time even after conditioning on choice set and covariates. 

\subsection{Identification of $m$ and $F^{\rumo}$}
Given the grand choice set $\mathcal{Y}$, the biggest possible support for $\rand{D}$ is $2^{\mathcal{Y}}\setminus\emptyset$.\footnote{$2^{\mathcal{Y}}$ denotes the set of all subsets of $\mathcal{Y}$.} Unfortunately, the information contained in $Y=\abs{\mathcal{Y}}$ outcomes over $T$ time periods may not be enough to identify the distribution supported on $2^Y-1=\abs{2^{\mathcal{Y}}\setminus\emptyset}$ points.\footnote{$\abs{\mathcal{Y}}$ denotes the cardinality of $\mathcal{Y}$.} Depending on the length of the panel $T$, to pin down the distribution of choice sets nonparametrically, we may need to assume that some subsets of $2^{\mathcal{Y}}\setminus\emptyset$ are never considered. That is, we may need to impose \emph{sparsity} assumptions.
\par
Let $K$ denote the biggest integer that is less than or equal to $(T-1)/2$. If $K\geq 1$, then we can take any subset of $\mathcal{T}=\{1,\dots,T\}$ of size $K$, $\mathcal{T}^*$, and define the corresponding subvector $\rand{y}\left(\mathcal{T}^*\right)=\left(\rand{y}_{t}\right)_{t\in \mathcal{T}^*}\in \mathcal{Y}^{K}$. Let $G$ be the conditional probability mass function of $\rand{y}\left(\mathcal{T}^*\right)$ conditional on $\rand{x}$ and $\rand{D}$. That is,
\[
G\left(y^{K}\mid D,x,\mathcal{T}^*\right)=\Prob{\rand{y}\left(\mathcal{T}^*\right)=y^{K}\mid\rand{D}=D,\rand{\x}=x}
\]
for all $y^K\in\mathcal{Y}^{K}$, $x\in\X$, and $D\in\mathcal{D}_x$. Note that under Assumption~\ref{ass: CIP}
\[
G\left(y^{K}\mid D,x,\mathcal{T}_i\right)=\prod_{t\in \mathcal{T}^*} F_t^{\rumo}\left(y_{t}\mid D,x\right)
\]
for any $y^K=\left(y_t\right)_{t\in\mathcal{T}^*}$, $x\in\X$, and $D\in\mathcal{D}_x$. In particular, when $K=1$, we have that $\mathcal{T}^*=\{t\}$ for some $t\in\mathcal{T}$ and $G\left(y^{1}\mid D,x,\mathcal{T}^*\right)=F_t^{\rumo}\left(y_{t}\mid D,x\right)$.
In general, $G$ captures the joint conditional distribution of $K$ different choices conditional on covariates and choice sets.
\par
We impose the following linear independence restriction on these conditional probability mass functions.

\begin{assumption}[Linear Independence]\label{ass:linearindependence}
For every $x\in X$ and $\mathcal{T}^*\subseteq\mathcal{T}$, $\abs{\mathcal{T}^*}=K$, the set $\left\{G(\cdot\mid D,x,\mathcal{T}^*)\right\}_{D\in\mathcal{D}_x}$ is a collection of linearly independent functions. That is, if
\[
\sum_{D\in\mathcal{D}_x}\alpha_DG(y^K\mid D,x,\mathcal{T}^*)=0
\]
for all $y^K$, then $\alpha_D=0$ for all $D\in\mathcal{D}_x$.
\end{assumption}
Note that $K\geq1$ if and only if $T\geq3$. Hence, $T\geq3$ is a \emph{necessary} condition for Assumption~\ref{ass:linearindependence} to be well-defined. 
\par
The rank conditions similar to Assumption~\ref{ass:linearindependence} are standard in the missclassification literature and the literature on finite mixtures (see, for instance, \citealp{hu2008identification, allman2009identifiability, kasahara2009nonparametric, an2010estimating, bonhomme2014nonparametric, kasahara2014non}). It essentially means that the variation in choice sets induces sufficient variation in the implied distributions over choices.\footnote{In the context of auctions, a similar assumption for $K=1$ has been made in \citet{an2017identification,mbakop2017identification}, and \citet{luo2018unobserved}.} We discuss Assumption~\ref{ass:linearindependence} in greater detail and provide sufficient conditions for it in the next section.
\par
We are ready to state our main result.
\begin{thm}\label{thm:identification}
Suppose Assumptions~\ref{ass:observables}-\ref{ass:linearindependence} hold. Then
$m(\cdot\mid x)$ and $\{F_t^{\rumo}(\cdot\mid D,x)\}_{t\in\mathcal{T}}$ are constructively identified for all $x\in X$ and $D\in\mathcal{D}_x$.
\end{thm}
Theorem~\ref{thm:identification} recovers nonparametrically the joint distribution of choice sets and choices. To the best of our knowledge, no other work on this topic achieves this. We emphasize that (i) we do not impose any structure on the statistical dependence between preferences and choice sets; (ii) we do not need to know the number of possible choice sets.
\par
Intuitively, if we observe a DM who over time chooses either of two alternatives (e.g., $\{y_1,y_2,y_1\}$), most likely, under Assumption~\ref{ass:regularity}, she considers these two alternatives. Another DM who chooses more alternatives (e.g., $\{y_1,y_2,y_3\}$), most likely has a bigger choice set than the previous DM. This variation in choices across time allows us to pin down the choice sets. After the choice sets are identified, the variation in choices within a given choice set (i.e. across DMs that have the same choice set) allows us to identify the distribution of choices conditional on choice sets.
\par
In other words, we have two competing forces in the model: consideration and preferences. Without the panel structure of the data, in general, it is hard to say whether a good is picked less often because it is rarely considered or because it is rarely picked when considered. However, with the panel data, if the good is picked by many DMs but these DMs do not pick it often over time, then we can conclude that this good is frequently considered, but on average is dominated by something else. In contrast, if a good is picked by small number of DMs, but these DMs choose it often across time periods, then the good is rarely considered, but if considered, then picked frequently.
\par
In one of the steps of the proof of Theorem~\ref{thm:identification}, we use the eigendecomposition argument of \citet{hu2008identification} and \citet{hu2013identification}. That is why we need to observe the choices of the same individual at least three times ($T\geq 3$). However, we do not need to impose any monotonicity restrictions on $F^{\rumo}$. Another difference from \citet{hu2008identification} and \citet{hu2013identification} is that we do not need to know the number of possible choice sets $d_{D,x}$. It can be identified from the data.\footnote{Also, in contrast to \citet{hu2013identification}, in general, we do not have a natural normalization for eigenvectors and cannot directly use them in our identification argument. Thus, we need to use a different argument to prove our result. See Appendix~\ref{app:proofs} for further details.} 

\subsection{Linear Independence and Choice Sets}\label{subsec: linindep}
To better understand Assumption~\ref{ass:linearindependence} consider the following simple examples. Suppose that $Y=\{1,2\}$ and $K=1$ (i.e., $T$ equals to 3 or 4). For a fixed $x$ and $\mathcal{T}^*=\{t\}$, the support of $\rand{D}$, $\mathcal{D}_x$, is a subset of $2^{\mathcal{Y}}\setminus\emptyset=\{\{1\},\{2\},\{1,2\}\}$. If we assume that $\mathcal{D}_x=\{\{1\},\{1,2\}\}$, then checking the linear independence condition is equivalent to checking whether the following matrix has full column rank:\footnote{The columns of this matrix correspond to different elements of $\mathcal{D}_x$. The rows correspond to different values $y^{K}$ can take.}
\begin{equation*}
\begin{tabular}{c|c|c}
$y^K \setminus D$&\{1\} &\{1,2\}\\
\cline{1-3}
$1$ & $1$ & $F_t^{\rumo}(1\mid \{1,2\},x)$\\ 
\cline{1-3}
$2$ & $0$ & $F_t^{\rumo}(2\mid \{1,2\},x)$\\ 
\cline{1-3}
\end{tabular}
\end{equation*}
This matrix has full column rank as long as Assumption~\ref{ass:regularity} is satisfied. Using similar argument, we can conclude that if $\abs{\mathcal{D}_x}\leq 2$, then Assumption~\ref{ass:linearindependence} is satisfied in this example. However, if $\mathcal{D}_x=\{\{1\},\{2\},\{1,2\}\}$, then the above matrix becomes
\begin{equation*}
\begin{tabular}{c|c|c|c}
$y^K \setminus D$&\{1\}&\{2\} &\{1,2\}\\
\cline{1-4}
$1$ & $1$ & $0$ & $F_t^{\rumo}(1\mid \{1,2\},x)$\\ 
\cline{1-4}
$2$ & $0$ & $1$ & $F_t^{\rumo}(2\mid \{1,2\},x)$\\ 
\cline{1-4}
\end{tabular}
\end{equation*}
and the linear independence condition fails to hold since the number of possible choice sets is bigger than the number of possible values $y^K$ can take.
If, next, we increase $K$ to $K=2$ (i.e. $T$ equals to 5 or 6), then for $\mathcal{T}^*=\{t_1,t_2\}$ the matrix becomes
\begin{equation*}
\begin{tabular}{c|c|c|c}
$y^K \setminus D$&\{1\}&\{2\} &\{1,2\}\\
\cline{1-4}
$(1,1)\tr$ & $1$ & $0$ & $F_{t_1}^{\rumo}(1\mid \{1,2\},x)F_{t_2}^{\rumo}(1\mid \{1,2\},x)$\\ 
\cline{1-4}
$(1,2)\tr$ & $0$ & $0$ & $F_{t_1}^{\rumo}(1\mid \{1,2\},x)F_{t_2}^{\rumo}(2\mid \{1,2\},x)$\\ 
\cline{1-4}
$(2,1)\tr$ & $0$ & $0$ & $F_{t_1}^{\rumo}(2\mid \{1,2\},x)F_{t_2}^{\rumo}(1\mid \{1,2\},x)$\\ 
\cline{1-4}
$(2,2)\tr$ & $0$ & $1$ & $F_{t_1}^{\rumo}(2\mid \{1,2\},x)F_{t_2}^{\rumo}(2\mid \{1,2\},x)$\\ 
\cline{1-4}
\end{tabular}
\end{equation*}
and Assumption~\ref{ass:linearindependence} is satisfied. As these examples demonstrate, for the linear independence condition to hold, we need to have either enough time periods or sparse $\mathcal{D}_x$ (i.e. relatively small support of $\rand{D}$). The former can be formalized as the following sufficient condition for Assumption~\ref{ass:linearindependence}.

\begin{prop}\label{prop: suff cond for linindep}
If Assumption~\ref{ass:regularity} holds and $K\geq Y$, then Assumption~\ref{ass:linearindependence} is satisfied.
\end{prop}

Proposition~\ref{prop: suff cond for linindep} provides a new, purely data-driven sufficient condition for the linear independence assumption. Proposition~\ref{prop: suff cond for linindep} is demanding in terms of observables -- we need to observe DMs for at least $2Y+1$ time periods. However, it does not impose \emph{any} restrictions on $\mathcal{D}_x$. For instance, $\mathcal{D}_x$ is allowed to include all nonempty subsets of $\mathcal{Y}$ (including singleton sets). To the best of our knowledge, this is the only identification result available in the literature that does not impose any restrictions on the random choice set. We note that having $d_{D,x}=\abs{\mathcal{D}_x}$ points in the support of $\rand{y}(\mathcal{T}^*)$ is a necessary condition for Assumption~\ref{ass:linearindependence} to hold. Thus, if $Y^K\geq 2^Y-1$, then this necessary condition is satisfied. Hence, even if $K<Y$, Assumption~\ref{ass:linearindependence} still may hold. 
\par

Now we investigate conditions that do not impose any restrictions on $T$, but restrict the support of choice sets by assuming a form of sparsity.
\begin{prop}\label{prop: suff cond for linindep2}
Suppose Assumption~\ref{ass:regularity} and one of the following conditions hold:
\begin{enumerate}
\item (Nestedness) For every $\x\in\X$, $\mathcal{D}_x$ is a collection of nested sets. That is, $\mathcal{\D}_\x=\{D_k\}_{k=1}^{d_{\D,\x}}$ such that $D_{k-1}\subseteq\D_{k}$ for $k=2,\dots,d_{\D,x}$.
\item (Excluded Choices) For every $\x\in\X$, $\mathcal{D}_x=\{D_k\}_{k=1}^{d_{D,x}}$ is such that for every $k$ there exists $y_k\in \mathcal{Y}$ such that $y_k\in D_{k}$, but $y_k\not\in D_{k'}$ for all $k'\neq k$.
\end{enumerate}
Then Assumption~\ref{ass:linearindependence} is satisfied.
\end{prop}

Conditions in Proposition~\ref{prop: suff cond for linindep2} are sufficient for Assumption~\ref{ass:linearindependence} to hold because they impose enough structure on the matrix constructed from $G$ to guarantee that it is of full rank. For instance, when $K=1$, nestedness implies that the matrix $[G\left(y\mid D,x,\mathcal{T}^*\right)]_{y\in \mathcal{Y}, D\in\mathcal{D}_x}=[F_t^{\rumo}(y\mid D,x)]_{y\in \mathcal{Y}, D\in\mathcal{D}_x}$ is triangular for any $t$.\footnote{We use $[a_{ij}]_{i\in I ,j\in J}$ to denote a matrix of the size $\abs{I}\times\abs{J}$ with entries of the form $a_{ij}$.} Next we give two examples of environments when conditions in Proposition~\ref{prop: suff cond for linindep2} may hold.

\begin{example}[Distance]\label{ex:distance}
Consider a DM who picks a restaurant at different time periods $t\in\mathcal{T}$. At every $t$ the preferences of the DM over restaurants are captured by $\rand{u}_t$. DMs live in the same location but are different in terms of the unobserved type of transportation they have access to. For instance, some DMs can only walk to restaurants, others can bike or take a cab. The realization of the random choice set $\rand{D}$ captures the set of restaurants the DM can get to. Condition (i) in Proposition~\ref{prop: suff cond for linindep2} is satisfied due to geographical nestedness: a person with a car can get to any place attainable by a bicyclist; a person with a bicycle can get to any place attainable by a pedestrian. 
\end{example}

\begin{example}[Brand/Product Loyalty]\label{ex:core}
Suppose $\mathcal{D}_x=\{\bar{D}\cup\{y_k\}\}_{k=1}^{d_{D,x}}$, where $y_k\neq y_{k'}$ for $k\neq k'$, and $y_k\not\in\bar{D}$ for all $k$. Then condition (ii) in Proposition~\ref{prop: suff cond for linindep2} is satisfied. Alternatives $y_k$ represent a particular brand or product that DM is loyal to. $\bar{D}$ can be thought of as the set of products considered by every DM (e.g., store brand). Brand or product loyalty is characterized by the fact that if DM considers $y_k$, she ignores all other options not present in $\bar{D}$. For instance, let $Y=\{1,2,3,4\}$ be a set of different brands of cereals, where $1$ represents a store brand. Then the choice sets $\{\{1,2\},\{1,3\},\{1,4\}\}$ represents three types of consumers that look at brands $2$, $3$, and $4$. However, everyone pays attention to the default store brand option $\bar{D}=\{1\}$. 
\end{example}

Note that Propositions~\ref{prop: suff cond for linindep} and~\ref{prop: suff cond for linindep2} do not assume that the identity or the number of support points $d_{D,x}$ is known or is the same for all $x$. Distinct in terms of covariates DMs may draw their choice sets from completely different distributions. This is empirically relevant since it allows us to analyze consumers from very heterogeneous markets.
\par
Propositions~\ref{prop: suff cond for linindep} and~\ref{prop: suff cond for linindep2} display two complementary and non-overlapping identifying features of the model. Proposition~\ref{prop: suff cond for linindep} states that if you have long enough panel, then no restrictions on the support of the choice sets are needed. In contrast, Proposition~\ref{prop: suff cond for linindep2} is not demanding in terms of data availability (i.e., $T$ can be as small as $3$), but requires more structure on the support of random consideration sets. We believe this trade-off is inevitable if one wants to achieve nonparametric identification of \emph{both} the distribution of unobserved choice sets and the distribution of choices conditional on choice set. 
\par
Importantly, Assumption~\ref{ass:linearindependence} can be satisfied in settings beyond the ones considered in Propositions~\ref{prop: suff cond for linindep} and~\ref{prop: suff cond for linindep2}. In particular, one of the implications of Assumption~\ref{ass:linearindependence} is a restriction on the cardinality of $\mathcal{D}_x$. Indeed, for $K=1$ (or $T=3$), it must be that for all $x\in X$, the number of points in the support of the random choice set, $d_{D,x}=\abs{\mathcal{D}_x}$, has to be smaller than or equal to the total number of alternatives, i.e., $d_{D,x}\leq Y$. In other words, the support of $\rand{D}$ needs to be sparse. In general, there is no need to impose a sparsity condition for identification of finite mixtures if the dependent variable is continuously distributed and the latent heterogeneity is discrete (e.g., \citealp{hu2013identification}). In our setting, the dependent variable has finite support, thus, we have to reduce the dimensionality of the problem by bounding the cardinality of the support of the latent choice sets. This sparsity restriction can be satisfied in many empirical settings. Suppose for every $x$
\[
\mathcal{D}_{x}=\{D\subseteq \mathcal{Y}\::\:\mathrm{L}_x\subseteq D\subseteq\mathrm{U}_x\},
\]
where $\mathrm{L}_x$ and $\mathrm{U}_x$ are some observed sets. Such a restriction on the choice sets appears in \citet{conlon2013demand, lu2014moment}, and \citet{gentry2016displays}. In this case, 
\[
d_{D,x}\leq\abs{\mathcal{Y}}\iff 2^{\abs{\mathrm{U}_x\setminus\mathrm{L}_x}}\leq Y.
\]
In the dataset used in \citet{conlon2013demand}, $2^{\abs{\mathrm{U}_x\setminus\mathrm{L}_x}}\leq 2^3=8$ (at most $3$ stock-out events) while total number of products considered is $44$. 

\subsection{Relaxing Conditional Independence Assumption}\label{sec: markov}
Assumption~\ref{ass: CIP} requires choices to be independent across $t$ conditional on the observed covariates and unobserved choice sets. In this section, we extend our analysis to environments with unobserved persistence over time. 
\par
We impose the following two restrictions on the dependence structure across $t$.
\begin{assumption}[Markovianity]\label{ass: markov}
For all $t$, $x$, $D$, and $y_1$,$y_2$, ...$y_{t}$
\[
\Prob{\rand{y}_t=y_t\mid \rand{y}_{t-1}=y_{t-1},\dots,\rand{y}_1=y_1,\rand{D}=D,\rand{x}=x}=\Prob{\rand{y}_t=y_t\mid \rand{y}_{t-1}=y_{t-1},\rand{D}=D,\rand{x}=x}.
\]
\end{assumption}

Assumption~\ref{ass: markov} is a standard markovianity assumption that requires future choices to be independent of the past choices as long as one conditions on the current choice. Although we assume the dependence on one lagged variable, the results in this section can be extended to cases with more lagged variables. However, the more lagged variables we allow, the longer panel we need to observe.\footnote{\citet{mbakop2017identification} uses markovianity of order statistics to identify the distribution of private valuations in auction settings.}

\begin{assumption}[Distribution Stability]\label{ass: dist stab}
For all $t,k$, $D$, $x$, and $y,y'$
\[
\Prob{\rand{y}_t=y\mid \rand{y}_{t-1}=y',\rand{D}=D,\rand{x}=x}=\Prob{\rand{y}_{k}=y\mid \rand{y}_{k-1}=y',\rand{D}=D,\rand{x}=x}.
\]
\end{assumption}

For all $y,y'\in \mathcal{Y}$, $x\in X$, and $D\in\mathcal{D}_x$ define
\[
F_d^{\rumo}(y\mid y',D,x)=\Prob{\rand{y}_T=y\mid \rand{y}_{T-1}=y',\rand{D}=D,\rand{x}=x}.
\]
Given Assumptions~\ref{ass: markov} and~\ref{ass: dist stab}, $F_d^{\rumo}$ is time invariant. 
Assumption~\ref{ass: dist stab} requires the conditional distribution of choices to be stable over time. It is only needed if one wants to extrapolate $F_d^{\rumo}$ to all time periods. Without Assumption~\ref{ass: dist stab} we can only identify the conditional distribution of choices conditional on the past choice and choice sets at one particular time period.
\par
In order to accommodate dynamics in our setting, we need to update the linear independence assumption (Assumption~\ref{ass:linearindependence}). Let $K_d$ denote the biggest integer that is less than or equal to $(T-3)/2$. If $K_d\geq 1$, then we can construct two nonoverlapping subsets of $\mathcal{T}$: $\mathcal{T}_{d,1}=\{1,\dots,K_d\}$ and $\mathcal{T}_{d,2}=\{K_d+2,\dots,2K_d+1\}$. Assume for a moment that $K_d=(T-3)/2$. Then $t=K_d+1$ separates $\mathcal{T}_{d,1}$ and $\mathcal{T}_{d,2}$; and $t=2K_d+2$ separates $\mathcal{T}_{d,2}$ from the last observation $\rand{y}_{T}$. For any $\mathcal{T}_{d,i}$, $i=1,2$, define $\rand{y}(\mathcal{T}_i)=(\rand{y}_{t})_{t\in \mathcal{T}_{d,i}}\in \mathcal{Y}^{K_d}$. In other words, we partition the sequence $\{\rand{y}_t\}_{t\in\mathcal{T}}$ into $5$ random vectors: $\rand{y}(\mathcal{T}_1),\rand{y}_{K_d+1},\rand{y}(\mathcal{T}_2),\rand{y}_{2K_d+2}$, and $\rand{y}_{T}$. 
The following lemma serves as the basis for our identification result in this section.
\begin{lem}\label{lem: dynam cond indep}
Under Assumption~\ref{ass: markov}, $\rand{y}(\mathcal{T}_1),\rand{y}(\mathcal{T}_2)$, and $\rand{y}_{2K_d+3}$ are conditionally independent conditional on $\rand{y}_{K_d+1}$, $\rand{y}_{2K_d+2}$, $\rand{x}$, and $\rand{D}$.
\end{lem}

Given the conditional independence we just need to reformulate the linear independence condition. Let $G_d$ be the conditional probability mass function of $\rand{y}(\mathcal{T}_{d,i})$ conditional on $\rand{y}_{K_d+1}$,$\rand{y}_{2K_d+2}$, $\rand{x}$, and $\rand{D}$. That is,
\[
G_d\left(y^{K_d}\mid y,y',D,x,\mathcal{T}_i\right)=\Prob{\rand{y}(\mathcal{T}_i)=y^{K_d}\mid \rand{y}_{K_d+1}=y,\rand{y}_{2K_d+2}=y',\rand{D}=D,\rand{\x}=x}
\]
for all $y^{K_d}\in\mathcal{Y}^{K_d}$, $y,y'\in\mathcal{Y}$, $x\in\X$, and $D\in\mathcal{D}_x$. 

\begin{assumption}[Dynamic Linear Independence]\label{ass: dynamic linearindependence}
For every $x\in X$, $y,y'\in\mathcal{Y}$, and $i\in\{1,2\}$, $\left\{G_d(\cdot\mid y,y',D,x,\mathcal{T}_i)\right\}_{D\in\mathcal{D}_x}$ is a collection of linearly independent functions.
\end{assumption}
Given the new linear independence condition we can formulate the following extension of Theorem~\ref{thm:identification} that allows for dependence of choices across time even after conditioning on choice sets and covariates.
\begin{prop}\label{prop: dynamic identification}
Suppose Assumptions~\ref{ass:observables},~\ref{ass:regularity},~\ref{ass: markov}-\ref{ass: dynamic linearindependence} hold. Then $m$ and $F_d^{\rumo}$ are constructively identified.
\end{prop}

\subsection{Variation in Covariates, Full Choice Set Variation, and Identification of $F^{\rumo}$}
Theorem~\ref{thm:identification} (or Proposition~\ref{prop: dynamic identification}) does not identify $F^{\rumo}(\cdot\mid D,x)=\{F_t^{\rumo}(\cdot\mid D,x)\}_{t\in \mathcal{T}}$ for all possible choice sets (i.e., for all $D\in 2^\mathcal{Y}\setminus\emptyset$). Instead, for a fixed value of the covariate $x$, it identifies $F^{\rumo}$ on the conditional support of the choice sets, $\mathcal{D}_{x}$.
To recover the random utility distribution over all possible choice sets and, thus, to uncover all possible substitution/complimentarity patterns we can strengthen our assumptions.\footnote{Two products may never appear in the same choice set. Hence, without extra assumptions we may not be able to figure out whether these products are compliments or substitutes.} We highlight that full choice set variation allows us to recover all possible ordinal information about the distribution of preferences in a fully nonparametric fashion. 
We underline that Theorem~\ref{thm:identification} does not use any variation in observed covariates $x$. In this section we use additional information contained in $x$ to learn more about choices of DMs.
\par
Suppose that the vector of covariates $x$ can be partitioned into $z\in Z$ and $w\in W$. Let $Z_w$ denote the conditional support of $z$ conditional on $\rand{w}=w$. 
\begin{assumption}[Excluded Covariates]\label{ass:excluded}
$F^{\rumo}(\cdot\mid \D,\x)=F^{\rumo}(\cdot\mid \D,\x')$ for all $x\neq x'$ with $\w=\w'$, for all $D$.
\end{assumption}
Assumption~\ref{ass:excluded} implies that there are covariates that can serve as exclusion restrictions: changes in the value of these covariates generates variation in choice sets, but does not affect the distribution of preferences. In many multinomial choice environments prices of goods can be taken as excluded covariates, since increasing prices shrink the set of feasible goods while not affecting preferences of DMs.\footnote{Prices affect indirect utility.} Another example is advertising: in many environments it is natural to assume that advertising increases the awareness about a product, but does not affect DM's preferences.  
\par
Define 
\[
\mathcal{B}_{w}=\bigcup_{\z\in\Z_\w}\mathcal{D}_{(\z\tr,\w\tr)\tr}
\]
for every $w\in W$. Note that when $w$ is fixed the preference distribution is also fixed. Thus, $\mathcal{B}_{w}$ contains all possible choice sets that DMs may face. The elements of $\mathcal{B}_{w}$ do not need to be sparse since Assumption~\ref{ass:linearindependence} needs to be satisfied after conditioning on both excluded and nonexcluded covariates. The variation in $\rand{z}$ can generate a substantial variation in the support of the random choice set
$\mathcal{D}_x$. For example, $\mathcal{B}_{w}$ can be equal to $2^{\mathcal{Y}}\setminus\emptyset$. 

\begin{cor}\label{cor:identification}
Under conditions of Theorem~\ref{thm:identification}, if 
\[
\mathcal{B}_{w}=2^{\mathcal{Y}}\setminus\emptyset,
\]
then $F^{\rumo}(\cdot\mid \D,x)$ is identified for all $x\in\X$ and $D\in2^{\mathcal{Y}}\setminus\emptyset$.
\end{cor}

\section{Estimation}\label{sec: estimation}
In this section we provide a computationally efficient and consistent estimator of $m$ and $F^{\rumo}$. First, we introduce some notation and objects that are necessary for estimation. Next, we discuss two straightforward consistent estimators and discuss their drawbacks -- one is computationally infeasible in moderate size problems, the other one (we call it the Step-1 estimator) is fast but does not perform well in finite samples. We then propose a new estimator that overcomes these two issues. This estimator regularizes the fast Step-1 estimator to achieve a better performance in finite samples.
\par
To simplify the exposition, we assume that $T=3$. Let $\mathrm{P}$ be the conditional probability mass function of choices conditional on covariates. That is,  
\[
\mathrm{P}(y_1,y_2,y_3\mid x)=\Prob{\rand{y}_1=y_1,\rand{y}_2=y_2,\rand{y}_3=y_3\mid \rand{x}=x}
\]
for every $y_1,y_2,y_3\in\mathcal{Y}$ and $x\in X$.
We assume that the analyst has access to a consistent estimator of $\mathrm{P}$, $\mathrm{\hat{P}}$. For instance, if the analyst observes a sample of size $n$ of i.i.d. observations coming from the joint distribution of $(\rand{y}_t)_{t\in\mathcal{T}}$ and $\rand{x}$, $\left\{\left(\rand{y}^{(i)}_t\right)_{t\in\mathcal{T}},\rand{x}^{(i)}\right\}_{i=1}^n$, and $X$ is a finite set, then for any $y_1,y_2,y_3\in\mathcal{Y}$ and $x\in X$ one can use
\[
\mathrm{\hat{P}}(y_1,y_2,y_3\mid x)=\dfrac{\dfrac{1}{n}\sum_{i=1}^n \Char{\rand{y}^{(i)}_1=y_1,\rand{y}^{(i)}_2=y_2,\rand{y}^{(i)}_3=y_3,\rand{x}^{(i)}=x}}{\dfrac{1}{n}\sum_{i=1}^n\Char{\rand{x}^{(i)}=x}}.
\]
For continuously distributed $x$, one can use any nonparametric estimator of a conditional expectation based on sieves or kernels (see \citealp{chen2007large} and \citealp{li2007nonparametric}).\footnote{For a recent application of a sieve estimator with continuous covariates see, for instance, \citet{kashaev2020identification}.} In our estimation routine, we take $\hat{P}$ as given.
\begin{assumption}
    There exists an estimator of $\mathrm{P}$, $\mathrm{\hat{P}}$, and a diverging sequence of positive natural numbers $\alpha_n$ such that
    $\alpha_n\left(\mathrm{\hat{P}}(y_1,y_2,y_3\mid x)-\mathrm{P}(y_1,y_2,y_3\mid x)\right)$ is stochastically bounded in probability for any $y_1,y_2,y_3\in\mathcal{Y}$ and $x\in X$. That is,
    \[
    \mathrm{\hat{P}}(y_1,y_2,y_3\mid x)-\mathrm{P}(y_1,y_2,y_3\mid x)=O_P(\alpha_n^{-1})
    \]
    for all $y_1,y_2,y_3\in\mathcal{Y}$ and $x\in X$.
\end{assumption}
The rate of convergence $\alpha_n$ depends on the asymptotic behavior of $\mathrm{\hat{P}}$. For instance, the estimator $\mathrm{\hat{P}}$ with discrete $\rand{x}$ above is $\sqrt{n}$-consistent estimator (i.e. $\alpha_n=\sqrt{n}$). 
\par
We are interested in estimating $F_t^{\rumo}(y\mid D,x)$ and $m(D\mid x)$ for all $y$, $D$, and $t$ for a given $x\in X$. We fix some $x$ and conduct the analysis below ``conditional on $\rand{x}=x$''. To simplify the exposition we drop $x$ from the notation (in our empirical application we estimate the model for all values of covariates).
\par
In the proof of Theorem~\ref{thm:identification}, we show that the cardinality of the support of $\rand{D}$ is equal to 
\[
d_{D}=\mathrm{rank}\left(\left[\sum_{k=1}^{Y}\mathrm{P}(i,j,k)\right]_{i,j\in \mathcal{Y}}\right),
\]
where $\mathrm{rank}(A)$ is the rank of matrix $A$. Thus,
given that $\mathrm{\hat{P}}$ converges in probability to $\mathrm{P}$, we can estimate the upper bound for the cardinality of $\mathcal{D}$ as
\[
\hat{d}_{D}=\mathrm{rank}\left(\left[\sum_{k=1}^{Y}\mathrm{\hat{P}}(i,j,k)\right]_{i,j\in \mathcal{Y}}\right).
\]
Instead of using this upper bound, we can infer the rank by applying any standard procedure (e.g., \citealp{cragg1997inferring}). However, since in the second step of our method, we pick no more than $\hat{d}_{D}$ sets that explain the data, asymptotically our procedure is still consistent.  

\subsection{Straightforward Estimator}
Given that our identification result implies that there is a unique distribution over choice sets $m$ and the distribution over choice conditional on choice sets $F^{\rumo}$, we can directly minimize the Euclidean distance to the observed distribution of choices over all possible combinations of $\hat{d}_\mathcal{D}$ subsets of $\mathcal{Y}$. The collection of subsets that minimizes this distance is a consistent estimator of the support of the choice sets.\footnote{Since there are finitely many collections of subsets (i.e., the parameter space is discrete), this estimator of the support of choice sets will converge arbitrary fast.} 
Formally, let $\Real_{+}^{a\times b}$ denote the space of matrices of size $a\times b$ with nonnegative real entries.  Fix a collection of $\hat{d}_\mathcal{D}$ different subsets of $\mathcal{Y}$ encoded by some matrix $\tilde{A}\in\Real_{+}^{Y\times \hat{d}_{D}}$, where
$\tilde{A}_{y,j}$ equals to $1$ if alternative $y$ belongs to a choice set $D_j$, and equals to $0$ otherwise. Then we can compute
\begin{align*}
    T(\tilde{A})=\min_{F^t\in\Real_{+}^{Y\times \hat{d}_{D}},\:M\in\Real_{+}^{\hat{d}_{D}}} &\sum_{y_1,y_2,y_3}\left\{\mathrm{\hat{P}}(y_1,y_2,y_3)-\left[\sum_{j=1}^{\hat{d}_{D}}F^1_{y_1,j}\cdot F^2_{y_2,j}\cdot F^3_{y_3,j}\cdot M_j\right]\right\}^2\\
    \text{s.t. }&\sum_{y\in\mathcal{Y}}F^t_{y,j}=1,\: j=1,2,\dots, \hat{d}_{D},\:t=1,2,3,\\
                & \sum_{j=1}^{\hat{d}_{D}}M_j=1,\\
                & F^t_{y,j}\leq \tilde{A}_{y,j}, \:y\in\mathcal{Y},\: j=1,2,\dots, \hat{d}_{D},\:t=1,2,3.
\end{align*}
The first and the second set of constrains require $F^t$ and $M$ to be proper probability distributions. The last set of constraints restricts the probability of choosing an alternative that is not considered (i.e., not in the choice set) to be zero.
\par
The collection $\tilde{A}^*$ that minimizes $T(\cdot)$ would deliver consistent estimator of the choice sets, and, thus, give us a consistent estimator of $F^{\rumo}$. Unfortunately, this procedure becomes computationally prohibitive for even relatively small $Y$. For instance, if, as in our empirical application, $Y=5$, then, without any restrictions on choices sets, there are $31!/(31-5)!>2\times10^7$ possible combinations of $5$ different sets out $31$ nonempty subsets of $\{1,2,\dots,5\}$. Even if every $T(A)$ is computed within $0.01$ sec, finding $\tilde{A}^*$ on a single core computer would take more than two days. In stark contrast, the procedure that we propose in the next section takes less than one minute on a single core computer.\footnote{Our simulations indicate that the estimation time of our procedure grows exponentially with $Y$. However, it is substantially faster than the procedure that checks all sets. For instance, our method takes about 6 hours to estimate a model with $Y=10$. The alternative would require solving $1023!/(1023-10)!>10^{18}$ optimization problems.} 
\par
The next estimator, which we call the Step-1 estimator, is similar to the previous estimator, but it does not force the constraints captured by matrix $\tilde{A}$. In particular, let the Step-1 estimator of $F^{\rumo}$ and $m$ be
\begin{align*}
    \bar{F}^*_{s1},\bar{M}_{s1}=&\argmin_{F^t\in\Real_{+}^{Y\times \hat{d}_{D}},\:M\in\Real_{+}^{\hat{d}_{D}}} \sum_{y_1,y_2,y_3}\left\{\mathrm{\hat{P}}(y_1,y_2,y_3)-\left[\sum_{j=1}^{\hat{d}_{D}}F^1_{y_1,j}\cdot F^2_{y_2,j}\cdot F^3_{y_3,j}\cdot M_j\right]\right\}^2\\
    \text{s.t. }&\sum_{y\in\mathcal{Y}}F^t_{y,j}=1,\: j=1,2,\dots, \hat{d}_{D},\:t=1,2,3,\\
                & \sum_{j=1}^{\hat{d}_{D}}M_j=1,
\end{align*}
where $\bar{F}_{s1}^*=(\bar{F}_{s1}^{1*},\bar{F}_{s2}^{2*},\bar{F}_{s3}^{3*})$.\footnote{Instead the described estimator, one can also use the estimator based on diagonalization argument as in \citet{hu2013identification}. Unfortunately, it suffers from the same issues in finite samples and performs a little worse in our simulations.}
\par
Given our identification result, the estimated $\bar{F}^{t*}_{s1,y,j}$ and $\bar{M}_{s1,j}$ are consistent estimators of $F_t^{\rumo}(y\mid D_j)$ and $m(D_j)$ since they are minimizing the Euclidean distance between the observed distribution and the distribution implied by the parameters.\footnote{If covariates are discrete, then instead of minimizing the Euclidean distance, one can also minimize the Kullback-Leibler divergence and obtain maximum-likelihood estimates.} However, because of the sampling uncertainty and numerical optimization errors the elements of matrix $\bar{F}_{s1}^{t*}$ that correspond to $F_t^{\rumo}(y\mid D)=0$ (i.e. $y\not\in D$) may not be exactly equal to zero. That is why, to recover the identity of choice sets, we trim the elements of $\bar{F}^{t*}_{s1,y,j}$ that are smaller than a prespecified $\varepsilon>0$. In our application and simulations, we set $\varepsilon=0.01$. Formally, we need the following strengthening of Assumption~\ref{ass:regularity}.
\begin{assumption}\label{ass:regularity2}
For every $\x\in\X$, $D\in\mathcal{D}_x$, $t\in\mathcal{T}$, and $y\in D$, 
\[
F_t^{\rumo}(\y\mid  D,\x)\geq\varepsilon
\]
for some known $\varepsilon>0$.
\end{assumption}
Thus, for every $y$ and $j$ we define the Step-1 estimator as
\[
F^{t}_{s1,y,j}=\dfrac{\bar{F}^{t*}_{s1,y,j}\Char{\bar{F}^{t*}_{s1,y,j}\geq\varepsilon}}{\sum_{y'=1}^Y\bar{F}^{t*}_{s1,y',j}\Char{\bar{F}^{s1,t*}_{y',j}\geq\varepsilon}}.
\]
Note that, instead of a fixed threshold, one can use a threshold $\varepsilon_n$ that converges to $0$ sufficiently slowly (e.g. $\varepsilon_n=\log(\log(n))/\sqrt{n}$ if $\alpha_n=\sqrt{n}$). In this case, Assumption~\ref{ass:regularity2} is not needed.\footnote{If some of the estimated sets appear more than one time (i.e., two columns of $\tilde{F}^{t}$ has the same zero components), then we can just drop one of them.}
\par
The Step-1 estimator does not require checking all possible collections of subsets of the grand choice set, however, it may perform poorly in finite samples (see Section~\ref{sec:simulations}). For instance, one problem of the Step-1 estimator is that it trims the unconstrained estimator of $F^{\rumo}$ to get the identity of consideration sets. This trimming, while delivering correct identities of the choices sets asymptotically, may be sensitive to the choice of $\varepsilon$ in finite samples. Thus, in finite samples, we propose to regularize the Step-1 estimator by using it as the starting point in the procedure described in the next section. 

\subsection*{Step-2 Estimator}

\noindent\textbf{Unconstrained Estimation.} Let $A\in\Real^{Y\times (2^Y-1)}$ be the matrix of zeros and ones that encodes all subsets of $\mathcal{Y}$. That is, $A_{y,j}=\Char{y\in D_j}$, $D_j\in 2^{\mathcal{Y}}\setminus{\emptyset}$. Let
\begin{align*}
    \bar{F}_{s2},\bar{M}_{s2}=&\argmin_{F^t\in\Real_{+}^{Y\times (2^{Y}-1)},\:M\in\Real_{+}^{2^{Y}-1}} \sum_{y_1,y_2,y_3}\left\{\mathrm{\hat{P}}(y_1,y_2,y_3)-\left[\sum_{j=1}^{\hat{d}_{D}}F^1_{y_1,j}\cdot F^2_{y_2,j}\cdot F^3_{y_3,j}\cdot M_j\right]\right\}^2\\
    \text{s.t. }&\sum_{y\in\mathcal{Y}}F^t_{y,j}=1,\: j=1,2,\dots, 2^Y-1,\:t=1,2,3,\\
                & \sum_{j=1}^{2^Y-1}M_j=1,\\
                & F^t_{y,j}\leq A_{y,j}, \:y\in\mathcal{Y},\: j=1,2,\dots, 2^Y-1,\:t=1,2,3.
\end{align*}
This optimization procedure is similar to the one in the previous section. However, it does not impose the sparsity condition -- all possible subsets of $\mathcal{Y}$ are allowed. As a result, this optimization problem, in general, may have several global minima since no assumptions on the number of choice sets are imposed. However, since the Step-1 estimator is consistent, there is a unique global minima to which the Step-1 estimator converges in probability. Hence, if we search for the optimum in the neighborhood of the Step-1 estimator, then the minimizer is still a consistent estimator of $m$ and $F^{\rumo}$. 
\par
\medskip
\noindent\textbf{Mixed Integer Optimization.} Note that in contrast to the Step-1 estimator, $\bar{F}_{s2}$ is forced to assign zeros at proper positions because of the constraints associated with matrix $A$. But, in finite samples, since no restrictions on the number of choice sets is imposed, it may assign positive mass to more that $\hat{d}_{D}$ sets. To solve this issue, we propose to solve the following mixed-integer problem: 
\begin{align*}
    \hat{B}_{s2},\tilde{M}_{s2}=&\argmin_{B\in\{0,1\}^{2^{Y}-1},\:M\in\Real_{+}^{2^{Y}-1}} \sum_{y_1,y_2,y_3}\left\{\mathrm{\hat{P}}(y_1,y_2,y_3)-\left[\sum_{j=1}^{2^{Y}-1}\bar{F}^1_{s2,y_1,j}\cdot \bar{F}^2_{s2,y_2,j}\cdot \bar{F}^3_{s2,y_3,j}\cdot M_{j}\right]\right\}^2\\
    \text{s.t. } & \sum_{j}^{2^Y-1}M_j=1,\\
                 & M_{j}\leq B_j,\: j=1,2,\dots, 2^Y-1,\\
                 & \sum_{j}^{2^Y-1}B_j\leq\hat{d}_{D}.
\end{align*}
Note that $B\in\{0,1\}^{2^{Y}-1}$ is a vector of zeros and ones and the objective function is similar to the least-squares objective since $\bar{F}_{s2}$ is fixed. Informally, one can think of $\bar{F}_{s2}$ as being a collection of regressors and $\sum_{j}^{2^Y-1}B_j\leq\hat{d}_{D}$ being a sparsity constraint: at most $\hat{d}_{D}$ regressors have to be active. As we discussed before, the model selection procedures, in general, are not consistent. But since we use a consistent estimator as a starting point in optimization the resulting $\hat{B}_{s2}$ correctly recovers the choice sets with probability approaching 1. Also since the last constraint is an inequality constraint, we may end up having less than $\hat{d}_{D}$ active choice sets.
\par
\medskip
\noindent\textbf{Final Step.} Finally, let $\hat{A}\in\Real_{+}^{Y\times \hat{d}_{D}}$ be the matrix of zeros and ones that encodes the choice sets estimated by $\hat{B}_{s2}$. Now, since we consistently estimated the choice sets on the previous step, we can finally estimate $F^{\rumo}$ and $m$:\footnote{Similar to the Step-1 estimator, instead of minimizing the Euclidean distance, we can conduct maximum likelihood estimation here when covariates are discrete.} 
\begin{align*}
    F_{s2},M_{s2}=&\argmin_{F^t\in\Real_{+}^{Y\times \hat{d}_{D}},\:M\in\Real_{+}^{\hat{d}_{D}}} \sum_{y_1,y_2,y_3}\left\{\mathrm{\hat{P}}(y_1,y_2,y_3)-\left[\sum_{j=1}^{\hat{d}_{D}}F^1_{y_1,j}F^2_{y_2,j}F^3_{y_3,j}M_j\right]\right\}^2\\
    \text{s.t. }&\sum_{y\in\mathcal{Y}}F^t_{y,j}=1,\: j=1,2,\dots, \hat{d}_{D},\:t=1,2,3,\\
                & \sum_{j=1}^{\hat{d}_{D}}M_j=1,\\
                & F^t_{y,j}\leq \hat{A}_{y,j}, \:y\in\mathcal{Y},\: j=1,2,\dots, \hat{d}_{D},\:t=1,2,3.
\end{align*}

\subsection{Discussions and Computational Aspects}
The key part of the proposed two-step estimator is recovering the choice sets (i.e., the support of $\rand{D}$). After the choice sets (or $\hat{B}_{s2}$) are found, in principle, any parametric or nonparametric estimation of $F^{\rumo}$ can be conducted (e.g., one can maximize the sample likelihood). 
\par
Our procedure is extremely fast and can be easily applied to choice sets of moderate size. For instance, in our empirical application with $Y=5$, our procedure completes the estimation in less than one minute on a single core computer. The main advantage is coming from employing mixed-integer programming to our problem. In particular, one can think of the search for $\hat{B}_{s2}$ as a regression problem with $2^Y-1$ ``regressors'' and at most $\hat{d}_\mathcal{D}$ nonzero coefficients. In the statistical literature, this problem is known as \emph{the best subset problem} (see \citealp{bertsimas2016best} and references therein for extensive discussion). Modern mixed-integer optimization algorithms can solve the best subset problem with thousands of observations and hundreds of active regressors within minutes.
\par
It is easy to impose any restrictions on the set of possible choice sets in our estimation procedure. For instance, if one wants to rule out singleton choice sets, one just needs to set the columns in matrix $A$ that correspond to singleton sets to zero columns and remove all repeated columns. Similarly, the lower and upper bound restrictions $\mathrm{L}_x$ and $\mathrm{U}_x$ discussed in Section~\ref{subsec: linindep} are easy to impose. If, for example, one wants to assume that, say, alternative $y$ is always considered, it is sufficient to set every element of the $y$-th row of $A$ to $1$ and again remove all repeated columns. 
\par
We conclude this section by noting that, given the discrete nature of the estimator of the latent choice set and the use of mixed-integer optimization, deriving confidence sets for the true choice sets, and, thus, for $m$, and $F^\rumo$ is nontrivial.\footnote{The existing methods conduct inference on parameters of interest under parametric assumptions about the distribution of choice sets. If these parametric assumptions are not valid, then the resulting standard error are incorrect.} However, if we assume that the choices sets are known, then the problem of estimation of $m$ and $F^\rumo$ is standard, and under the standard regularity conditions, one can conduct inference either by using normal approximations or bootstrap depending on the way one estimates $\mathrm{P}$.\footnote{In general, asymptotic properties of estimated $m$ and $F^{\rumo}$ depend on whether one has continuous covariates and uses kernels or sieves.} We leave the problem of constructing confidence sets for model parameters when the choice sets are also estimated for future work.

\section{Finite-sample Performance of the Estimator}\label{sec:simulations}
This section aims to analyze the finite sample performance of the estimator we propose in Section~\ref{sec: estimation}. Since the main innovation of our procedure is the estimation of choice sets, here we only present results for choice sets estimation. For results concerning estimation of $m$ and $F^{\rumo}$ see Appendix~\ref{app: MC simulations}.
\par
First, we define the data generating processes (DGPs) used in simulations. In all experiments we assume that there are no covariates, there are $T=3$ time periods, $Y=5$ alternatives, $d_{\mathcal{D}}=5$ choice sets, and $F^{\rumo}_1=F^{\rumo}_2=F^{\rumo}_3$. Every DGP is characterized by two matrices: $Pyd\in\Real^{5\times5}$ and $Pd\in\Real^5$.  $Pyd$ and $Pd$ is such that $Pyd_{y,j}=\Prob{\rand{y}_t=y\mid D_j}$ and $Pd_j=\Prob{\rand{D}=D_j}$. In other words, every column of $Pyd$ corresponds to a choice set. For instance, the fourth column of $Pyd$ in DGP1 indicates that the fourth choice set has two elements, $\{1,4\}$, and that conditional on $\rand{D}=\{1,4\}$ alternative $1$ is picked with probability $0.4$. Since the fifth element of $Pd$ is $0.15$, the probability of considering $\{1,4\}$ is $0.15$.
\par
\noindent\textbf{DGP1:} 
\[
Pyd=\left(\begin{array}{ccccc}
    1& 0.6& 0.5& 0.4& 0.2\\
    0& 0.4& 0& 0& 0\\
    0& 0& 0.5& 0& 0\\
    0& 0& 0& 0.6& 0\\
    0& 0& 0& 0& 0.8
\end{array}\right),\quad 
Pd=\left(\begin{array}{c}
    0.2\\
    0.15\\
    0.3\\
    0.15\\
    0.2
\end{array}\right).
\]
\noindent\textbf{DGP2:} 
\[
Pyd=\left(\begin{array}{ccccc}
    1& 0.6& 0.5& 0.25& 0.1\\
    0& 0.4& 0.2& 0.35& 0.25\\
    0& 0& 0.3& 0.25& 0.15\\
    0& 0& 0& 0.15& 0.3\\
    0& 0& 0& 0& 0.2
\end{array}\right),\quad 
Pd=\left(\begin{array}{c}
    0.2\\
    0.15\\
    0.3\\
    0.15\\
    0.2
\end{array}\right).
\]
The cardinality of the choice sets in DGP1 does no vary much. It assigns positive probability to the choice sets of cardinality less than 3 only. For example, alternative 1 enters all choice sets.\footnote{We impose this restriction in the estimation step.} All other options enter only one choice set.  DGP2 is very heterogeneous in terms of the size of the choice sets. The smallest one contains only alternative 1. The biggest one contains all alternatives. Note that these DGPs satisfy conditions of Proposition~\ref{prop: suff cond for linindep2} (Excluded Choices and Nestedness).
\par
We evaluate the ability of the Step-1 and Step-2 estimators to correctly recover sets. The results are presented in Table~\ref{table: correct sets percent}. As expected, the performance of both estimators improves with the sample size. The Step-2 estimator outperforms the Step-1 estimator in all experiments. The gains from the Step-2 estimator are especially striking for DGP2. For example, for a sample size of $2000$ the Step-1 estimator correctly recovers all 5 sets only in 4 percent of cases. However, the Step-2 estimator recovers all 5 sets in 29 percent of cases.
\begin{table}[h]
\centering
\begin{threeparttable}
\centering
\caption{Percent of Correctly Estimated Sets}\label{table: correct sets percent}
\begin{tabular}{lccccc}
\hline
\hline
Sample Size&    & 2000 & 5000 & 10000 & 50000\\ 
\hline
DGP1& Step-1                & 63.0 & 72.6 & 80.2  & 95.8\\
    & Step-2                & 64.1 & 86.8 & 93.9  & 99.9\\
DGP2& Step-1                & 4  & 9.9 & 24.8  & 77.6\\
    & Step-2                & 29 & 45.7 & 63.4  & 93.5\\
\hline
\end{tabular}
\vspace{1ex}
\begin{tablenotes}
\item {\footnotesize Notes: Number of replications=1000, $\varepsilon=0.01$. Results are rounded to 1 digit.}
\end{tablenotes}
\end{threeparttable}
\end{table}

To see how noisy the estimates of choice sets can be we also compute the average number of correctly recovered sets. As Table~\ref{table: average correct sets} shows, our estimator on average finds correctly more than 4 out of 5 sets. The Step-1 estimator recovers at least 3 out of 5 sets correctly on average. The performance of both estimators improves with the sample size.

\begin{table}[h]
\centering
\begin{threeparttable}
\centering
\caption{Average Number of Correctly Estimated Sets}\label{table: average correct sets}
\begin{tabular}{lccccc}
\hline
\hline
Sample Size&    & 2000 & 5000 & 10000 & 50000\\ 
\hline
DGP1& Step-1                & 4.63 & 4.73 & 4.8  & 4.96\\
    & Step-2                & 4.60 & 4.86 & 4.94  & 5\\
DGP2& Step-1                & 3.33  & 3.56 & 3.92  & 4.77\\
    & Step-2                & 4.14 & 4.39 & 4.61  & 4.93\\
\hline
\end{tabular}
\vspace{1ex}
\begin{tablenotes}
\item {\footnotesize Notes: Number of replications=1000, $\varepsilon=0.01$, number of sets=5. Results are rounded to 2 digits.}
\end{tablenotes}
\end{threeparttable}
\end{table}

\section{Empirical Application: Brand Choice Set Variation and Price Elasticity in the Ready-to-Eat Cereal Market.}\label{sec: empirical application}
In this section, we study the effects of brand choice set variation at the market level on consumption of the RTE cereal market using the Nielsen Homescan Panel (Homescan). The RTE cereal industry has been previously analyzed under the assumption that DMs consider all available brands (e.g.,  \citealp{nevo2000practitioner,nevo2001measuring}). We want to analyze to what extent this assumption is valid and what implications it has on parameters of interest such as price elasticity. (However, our results are not fully comparable to these previous results since we have a different dataset with richer variation needed in this setup.) The RTE cereal market is known to be highly concentrated, with high differentiation, large advertisement-to-sale ratios, and product innovation \citep{nevo2001measuring}. All of these factors suggests high variability in choice sets because of consumer loyalty, geographical variation in product availability, and targeted advertisement campaigns. We confirm this insight in our quantitative analysis and uncover substantial choice set variation. Moreover, we show that ignoring this latent choice set heterogeneity leads to biased estimates of price elasticities in a simple model of demand.
\par
First, we describe the dataset and empirical validity of our assumptions. Next, we apply our estimator to obtain the unobserved distribution of the choice sets and the distribution of choices given the choice sets. Finally, in the spirit of \citet{berry1995automobile,nevo2000practitioner, nevo2001measuring}, we estimate a simple parametric model of the RTE cereal demand. 

\subsection*{Data Construction}
We consider $5$ brands of RTE cereal: Store brand (CTL), General Mills (GM), Kellogg (K), Quaker (Q), and other brands of RTE cereal (O). We record only purchases of households that buy $1$ brand per trip.\footnote{The households that buy more than $1$ brand in a given trip are dropped from the sample to avoid dealing with bundling.} We focus on households that are frequent buyers. We define frequent buyers as households that buy at least one RTE cereal in $3$ consecutive trips.\footnote{A trip is an instance of a household member going to a store and purchasing at least one item that is recorded in the Homescan.} The majority of households in our sample makes $1$ trip per week. Thus, the predominant time frequency of our dataset is weekly. We focus on trips and households present in the Homescan in 2016-2018.\footnote{We eliminate from our sample trips happening in December and January, because of their strong seasonality effects on RTE cereal consumption.} We include only the $3$ earliest consecutive trips per household. Each household appears only once in the cross-section.\footnote{To ensure this we consider the first $3$ trips per year per household. Then we create a unified panel with the information of years $2016-2018$, and we balance the panel keeping only the first $3$ trips. Hence, if any household appears in all three years, we keep only its $2016$ observations.} We consider a balanced panel by dropping any household that does not have $3$ consecutive trips in a given year. We end up having $T=3$ consecutive choices of $47,509$ households.
\par
There are only $2$ product characteristics available in the Homescan and Nielsen Retail Scanner: price of a unit (USD) and size of it (ounces).\footnote{We also know barcodes of every purchase. Unfortunately, it is hard to match these barcodes with actual products to obtain additional product characteristics since these barcodes change over time and some products are not produced anymore.} The dataset also contains information on zip-codes for every household/purchase in the sample. We use the Nielsen Retail Scanner and the Homescan to construct the dataset on prices and sizes by pooling all the information on prices per UPC code (barcode) of all RTE cereals by week and location (3 digit zip-code).\footnote{To obtain prices in the Homescan we use the paid price (including discounts) divided by the number of units, and we drop from our sample those that pay a zero price after discount.} Then we compute the mean price of every brand at every location.\footnote{We average across weeks to diminish measurement error in prices and because there are some missing prices per brand.} Brand-location size variable is built similarly.
As a result, given the information on the location of every household, we match every purchase with the price and size.
\par
Despite having a relatively large sample, there are too many 3-digit level zip-codes to treat them as markets. Thus, to increase the number of observations per market, similar to \citet{nevo2001measuring}, we use prices and geographic coordinates (i.e. longitude and latitude) of every location to define markets. In particular, we define a market by employing K-means clustering with the Euclidean norm using centroids based on prices and geographic location. In other words, we group together households that live close to each other and face similar prices. We initialize the K-means and fix the number of markets using the 3-digit zip-code. All locations with less than $2000$ households are collapsed to a single dummy location.\footnote{This quantity was chosen on the basis of simulations, to ensure a sufficiently high number of observations per market.} In total we obtain $34$ markets (see Figure~\ref{Fig: set map}).\footnote{We obtain qualitatively the same results when we increased the number of markets to $72$.} Finally, we aggregate prices on the market-brand level. 
\par
Since the dataset contains information on the household's income, the age of the household's head, and the size of the household, we also compute the average (on the market level) income, the age of the head of the household, and the household size. We use these demographics in our analysis of own-price elasticity. Summary statistics for demographic variables are provided in Table~\ref{table: summary 1}

\begin{table}[h]
\centering
\begin{threeparttable}
\centering
\caption{Summary Statistics of Demographic Variables}\label{table: summary 1}
\begin{tabular}{lccccc}
\hline
\hline
Variable                    & Mean      & Median     & Std    & Min    & Max  \\ 
Average Age (years)              & 54.33   & 54.27 & 1.55 & 49,87 & 57.12\\
Average Income (USD)                 & 23,333  & 22,503 & 4,506.25 & 14,543.4 & 32,580\\
Average HH Size                & 2.7  & 2.7 & 0.12 & 2.49 & 3.18 \\
\hline
\end{tabular}
\vspace{1ex}
\begin{tablenotes}
\item {\footnotesize Notes: These summary statistics are computed for 34 markets. For instance, the minimum market average age is the smallest among 34 markets market-average age, not the age of youngest head of the household in the sample.}
\end{tablenotes}
\end{threeparttable}
\end{table}
\par
Since we mainly focus on frequent buyers (i.e., weekly purchases) of RTE cereals, we believe that the assumptions that choice sets are stable across time and preferences are conditionally independent across time are satisfied. It is less likely that the choice set changes within a short time horizon\footnote{For instance, product loyalty can be thought as a special case of system-1 thinking \citep{kahneman2003perspective}, hence, it is revised less often than utility maximization. Also, any product innovation within brands may take more time than three weeks which is the modal time window in our dataset.} and there are unobserved shocks to preferences over cereals that are correlated across time. That is, we believe that after controlling for observed characteristics and choice sets, all variation in choices is driven by idiosyncratic taste shocks.\footnote{In our application we use an additive random utility framework where the mean utility is assumed to be stable in the time-window, but taste shocks are idiosyncratic. Arguably, in a short time-window it is less likely that the DM adapts her mean utility due to structural environmental changes. In addition, \citet{saito2020repeated} show that the continuation problem can be ignored if preferences are separable and additive in time, which is a standard assumption in applied work.} Moreover, both assumptions (or their stronger versions) are usually made in the literature on estimation of demand systems using individual and market level data. For instance, \citet{nevo2001measuring} assumes independence of measurements across markets, where markets are defined as a pair of location and time window, and known fixed choice sets.

\subsection*{Nonparametric Estimation of Consideration Sets and True Market Shares}
We estimate $m$ and $F^{\rumo}$ for every market. Let $\hat{m}(D|j)$ denote the estimated probability that set $D$ is considered in market $j$. Since after averaging there is no within market variation, conditioning on the market also conditions on the product and consumers characteristics on top of unobserved market fixed effects. 
\par
Using the estimated $\hat{m}$, first we find that \emph{all} markets have less than 5 sets that are considered by more than $10$ percent of households in the market. That is, 
\[
\sum_{j}\Char{\sum_{D}\Char{\hat{m}(D|j)>0.1}\geq 5}=0.
\]
Even if we lower the threshold to $5$ percent, more than $20$ percent of markets have less than 5 choice sets. That is, among $5$ estimated sets at least $1$ set is considered by less than $10$ percent of population in every market, and a sizable fraction of markets have at least one set that is faced by less than 5 percent of consumers. These findings lend support to our sparsity assumption.
\par
\begin{figure}
		\centering
		\includegraphics[width=0.5\textwidth]{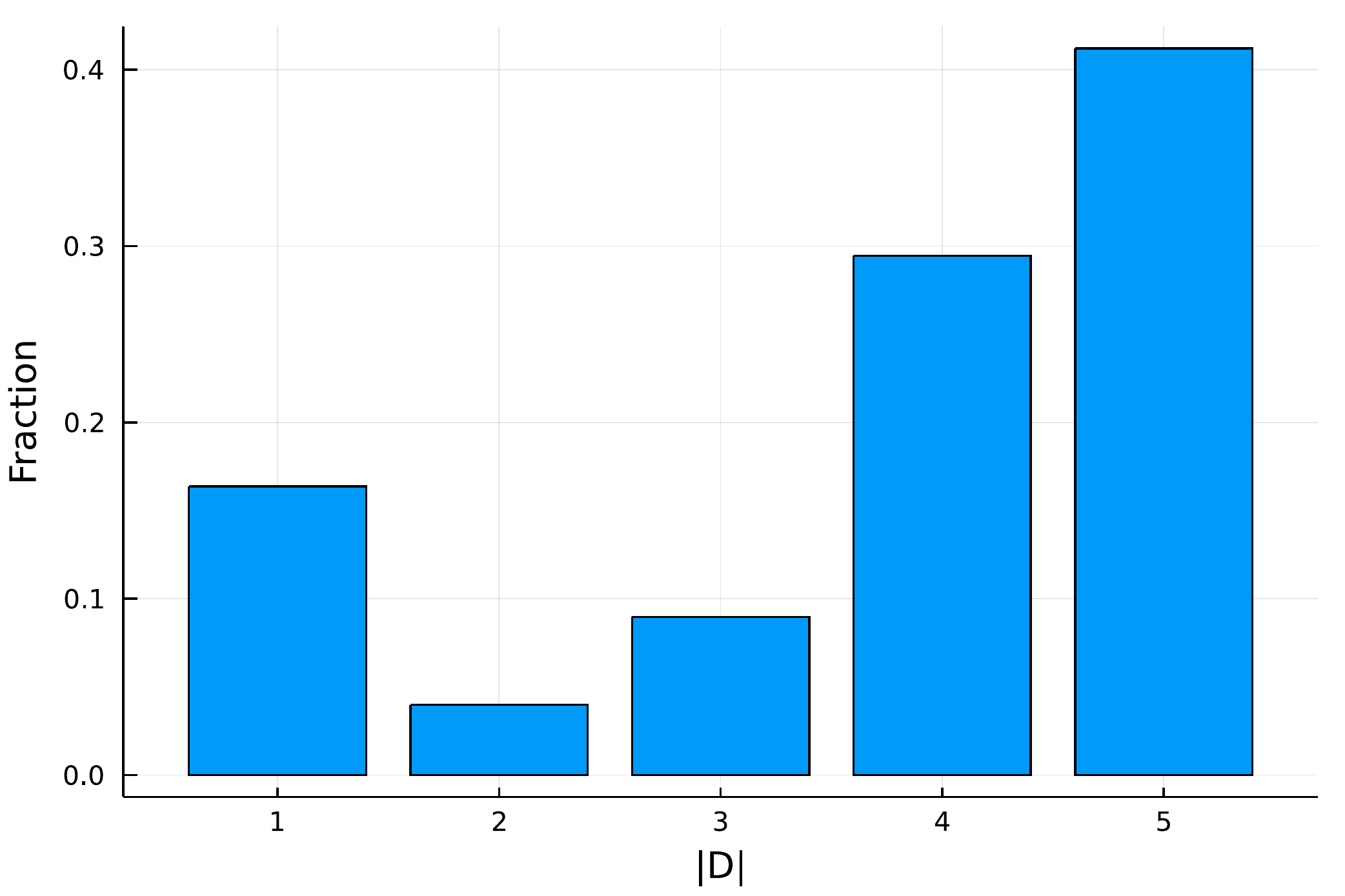}
	\caption{\textsc{Proportion of Individuals Considering Sets of Given Cardinality}: $|D|$ denotes the size of the choice set. }\label{Fig: average m cardinality}
\end{figure}

Next, we compute the estimated proportion of individuals in the sample who considered sets of a given cardinality $l$
\[
\sum_{j,D}\Char{\abs{D}=l}\hat{m}(D|j)w_j,
\]
where $w_j=N_j/N$ is a fraction of the whole sample (of size $N$) that is coming from market $j$ (of size $N_j$). As Figure~\ref{Fig: average m cardinality} demonstrates, sets of all sizes are considered. The vast majority (about $70$ percent) of the sample considered sets of cardinality $4$ and $5$. Given that most likely all 5 brands are usually present, these individuals can be thought of as full consideration individuals that we usually work with in discrete choice settings. However, about $16$ percent of DMs only considered one brand. These are super loyal consumers that always purchase the same brand no matter what. 
\par
\begin{figure}
		\centering
		\includegraphics[width=0.7\textwidth]{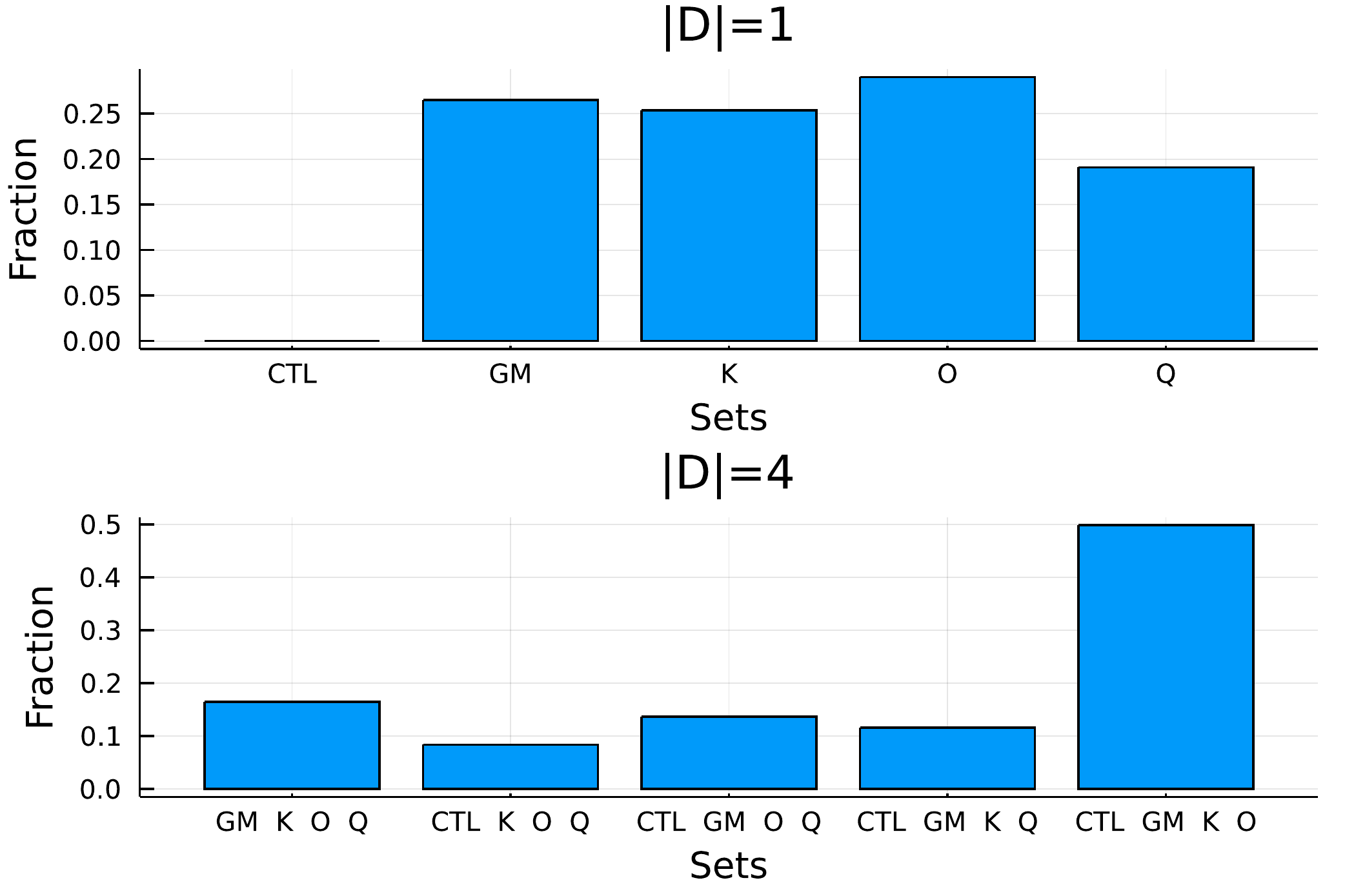}
	\caption{\textsc{distribution of Sets of Given Cardinality}: $|D|$ denotes the size of the choice set. CTL=Store brand, GM=General Mills, K=Kellogg, Q=Quaker , O=other brands. }\label{Fig: average m 1 and 4}
\end{figure}

Next, we consider the composition of sets of cardinality $1$ and $4$ (there is only one set of size $5$). The results are presented in Figure~\ref{Fig: average m 1 and 4}. Interestingly, Store brand (CTL) is never considered alone. The rest of the brands are almost equally likely considered by those who only look at one brand (Quaker has the smallest share of about $19$ percent). Among those who considered sets of size $4$, almost half considered everything but Quaker. The rest of sets of cardinality $4$ have similar shares.  
\par
\begin{figure}
		\centering
		\includegraphics[width=0.5\textwidth]{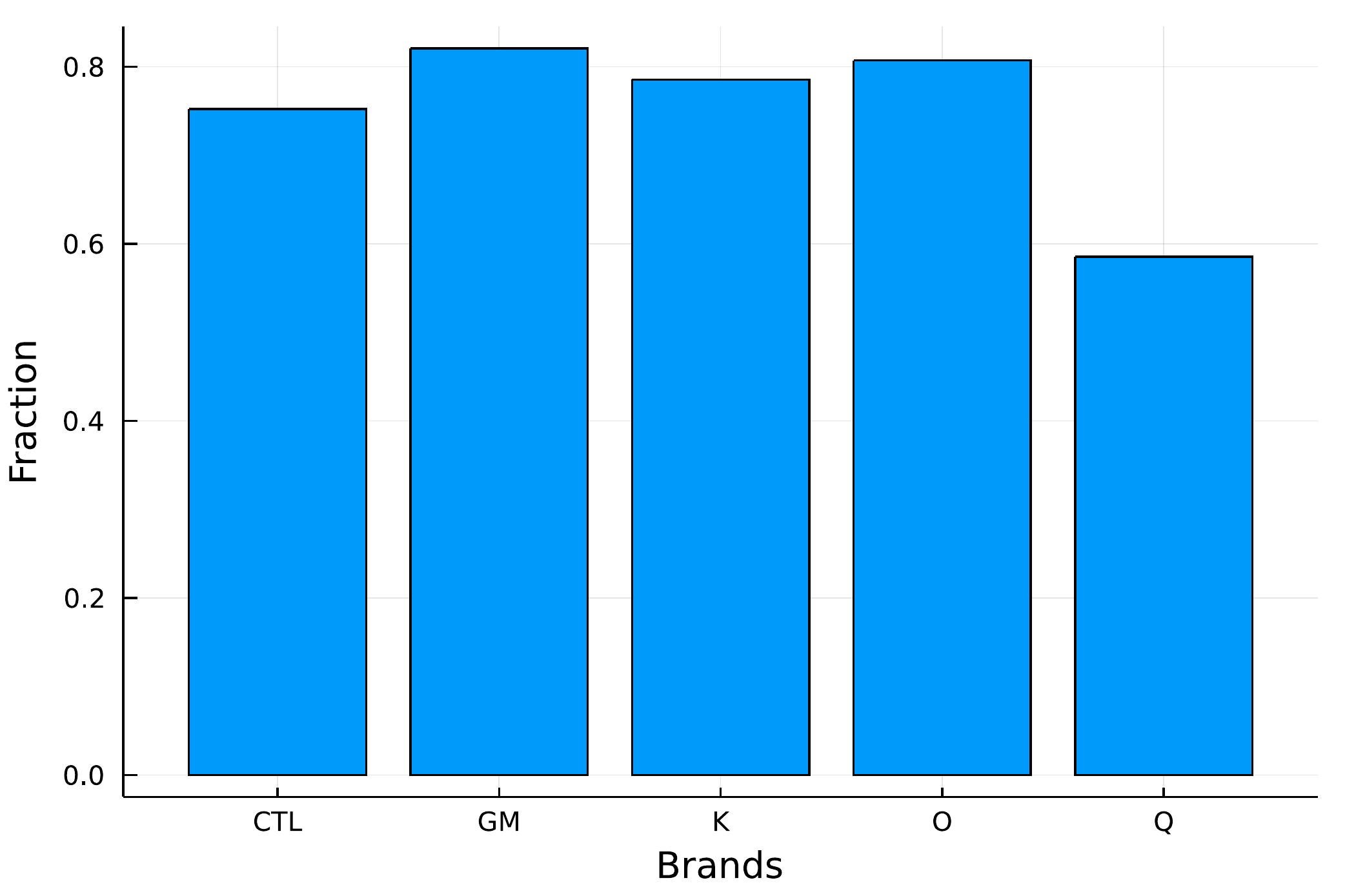}
	\caption{\textsc{Proportion of Individuals Considering a Brand}: CTL=Store brand, GM=General Mills, K=Kellogg, Q=Quaker , O=other brands. }\label{Fig: average m brand}
\end{figure}

Next, we compute the fraction of DMs who paid attention to a set that contains a given brand $b$
\[
\sum_{j,D}\Char{b\in D}\hat{m}(D|j)w_j.
\]
Similar to Figure~\ref{Fig: average m 1 and 4}, Figure~\ref{Fig: average m brand} indicates that Quaker is considered less often (about 58 percent of DMs) than other brands (about 80 percent of DMs).
\par
\begin{figure}
		\centering
		\includegraphics[width=1.0\textwidth]{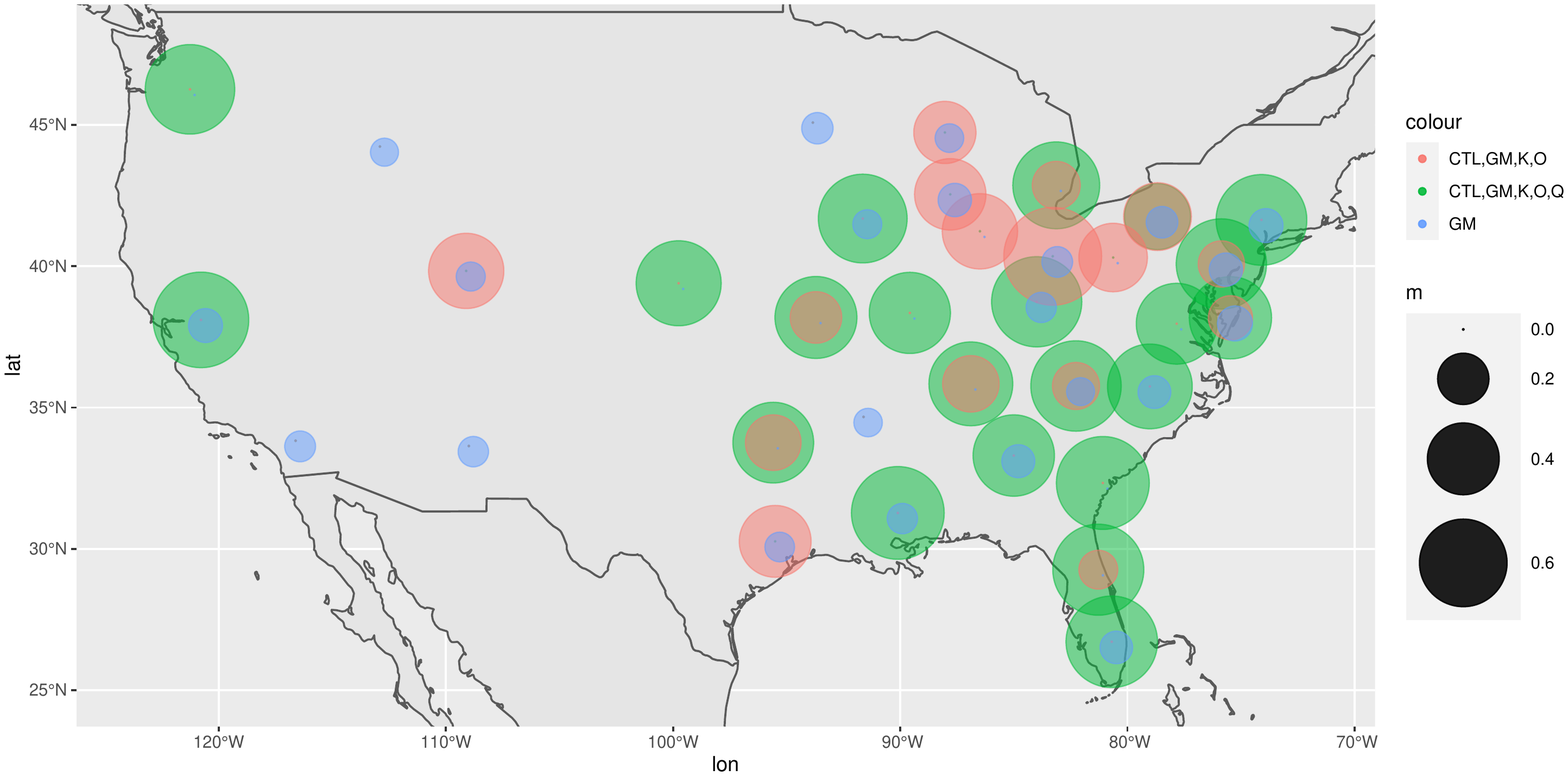}
	\caption{\textsc{Location of Markets and Consideration Probabilities of Some Choice Sets}: CTL=Store brand, GM=General Mills, K=Kellogg, Q=Quaker , O=other brands. The size of every circle corresponds to the estimates of consideration probability $\hat{m}$. The color of every circle corresponds to a different choice set: Red=\{CTL,GM,K,O\}, Green=\{CTL,GM,K,O,Q\}, Blue=\{GM\}.  }\label{Fig: set map}
\end{figure}

Finally, there are just 2 sets that attract more than 5 percent of DMs: the set that contains all $5$ brands (about 40 percent of DMs) and the set that contains all brands but Quaker (about 15 percent). Figure~\ref{Fig: set map} depicts the geographical location of our markets and the fractions of the population in these markets that consider the two most frequent sets. We also add those who only consider General Mills alone for a better glimpse of the heterogeneity in choice sets.
\par
Overall, we can conclude that although most DMs seem to consider almost all brands, there is a sizeable fraction of those who only consider one brand. Moreover, Quaker is considered less often than other brands and Store brand is always considered with other alternatives.
\par
Next, we consider the estimates of $F^{\rumo}$ per market and consideration set. To simplify the exposition, we focus on the 5 biggest (in terms of numbers of observations )markets ($j\in\{1,2,13,16,18\}$) and $t=1$.\footnote{The results for $t=2$ and $t=3$ are qualitatively the same.} Tables~\ref{table: naive shares} and~\ref{table: true shares} display estimated market shares of all brands assuming that every DM considered all brands (i.e. model with full consideration) and estimated market shares obtained via our procedure, respectively. The largest difference between the two shares is for Kellogg in market $16$. It is about 33 percentage points or 200 percent. Interestingly, the results of our estimations imply that in market $18$ there were no DMs who considered all 5 brands. If we assume that rankings of shares correspond to mean utilities of brands (i.e. additive random utility), then Table~\ref{table: naive shares} implies that, on average, General Mills is the most preferred brand in 4 out of 5 biggest markets if one assumes full consideration. If one allows for unobserved choice sets, then General Mills is the best alternative only in market 16. Kellogg is the best brand in markets 1 and 2. 

\begin{table}[h]
\centering
\begin{threeparttable}
\centering
\caption{Full Consideration Market Shares Assuming Observed Choice Sets}\label{table: naive shares}
\begin{tabular}{lccccc}
\hline
\hline
Brand/Market    & 1      & 2     & 13    & 16    & 18  \\ 
CTL             & 0.15   & 0.152 & 0.207 & 0.161 & 0.123\\
GM              & 0.316  & 0.293 & 0.269 & 0.311 & 0.292\\
K               & 0.283  & 0.286 & 0.293 & 0.279 & 0.26 \\
O               & 0.183  & 0.219 & 0.169 & 0.186 & 0.249\\
Q               & 0.069  & 0.05  & 0.062 & 0.063 & 0.076\\
\hline
\end{tabular}
\vspace{1ex}
\begin{tablenotes}
\item {\footnotesize Notes: Results are rounded to 3 digits.}
\end{tablenotes}
\end{threeparttable}
\end{table}

\begin{table}[h]
\centering
\begin{threeparttable}
\centering
\caption{Full Consideration Market Shares Assuming Unobserved Choice Sets}\label{table: true shares}
\begin{tabular}{lccccc}
\hline
\hline
Brand/Market    & 1      & 2     & 13    & 16    & 18  \\ 
CTL             & 0.1    & 0.099 & 0.123 & 0.117 & n/a\\
GM              & 0.334  & 0.319 & 0.219 & 0.372 & n/a\\
K               & 0.346  & 0.32  & 0.097 & 0.226 & n/a\\
O               & 0.132  & 0.202 & 0.497 & 0.196 & n/a\\
Q               & 0.088  & 0.061 & 0.064 & 0.089 & n/a\\
\hline
\end{tabular}
\vspace{1ex}
\begin{tablenotes}
\item {\footnotesize Notes: Results are rounded to 3 digits.}
\end{tablenotes}
\end{threeparttable}
\end{table}
Moreover, different in terms of choice sets DMs display different preferences over brands. For instance, in market 1, those who consider all 5 brands prefer Kellogg over all other brands. At the same time, those who do not consider Quaker predominantly buy other brands of cereals (see Table~\ref{table: shares market 1}). This emphasizes the importance of allowing for correlation between preferences and choice set when estimating the model.      
\begin{table}[h]
\centering
\begin{threeparttable}
\centering
\caption{Market Shares for Two Choice Sets in Market 1}\label{table: shares market 1}
\begin{tabular}{lcc}
\hline
\hline
Brand/Set       & \{CTL,GM,L,Q,O\}      & \{CTL,GM,L,O\}\\ 
CTL             & 0.1                   & 0.081\\
GM              & 0.334                 & 0.161\\
K               & 0.346                 & 0.11\\
O               & 0.132                 & 0.648\\
Q               & 0.088                 & 0.0\\
\hline
\end{tabular}
\vspace{1ex}
\begin{tablenotes}
\item {\footnotesize Notes: Results are rounded to 3 digits.}
\end{tablenotes}
\end{threeparttable}
\end{table}
Overall, our estimates suggest that allowing for unobserved choice sets affects not only point estimates of market shares, but also rankings of brands. Moreover, DMs that consider different choice sets may have different preferences.

\subsection*{Parametric Estimation of Price Elasticity with Hidden Choice Set Variation}
Note that the estimated $F^{\rumo}$ for every market is a collection of market shares of every alternative in a given market for different groups of consumers. For instance, $\rand{s}_{y,D,j,t}=\hat{F}_t^{\rumo}(y|D,j)$ is the estimated market share of brand $y$ among those consumers that face the choice set $D$ at time $t$ in market $j$ and decided to purchase something.\footnote{Our DMs are RTE cereal frequent buyers. Hence, we do not allow for the option of not buying anything.} So, we can proceed as if the estimated market shares are the true market shares and parametrize $F^{\rumo}$.\footnote{The market shares computed directly from the data are not the true market shares but rather a mixture of the market shares from different choice sets.} In our application, we take the standard logit specification of $F^{\rumo}$. In particular, following \citet{nevo2001measuring}, we assume that the random utility that consumer $i$ gets from brand $y$ in choice set $D$ at time $t$ in market $j$ is
\[
\alpha_{y,D}+\beta_D\rand{p}_{y,j}+\rand{r}_{j}\tr\gamma_{y,D}+\xi_{y,D}+\Delta\rands{\xi}_{y,D,j,t}+\rands{\varepsilon}_{i,y,D,t},
\]
where $\rand{p}_{y,j}$ is the average market price of brand $y$ in market $j$; $\rand{r}_{j}$ is the vector of market demographics that consists of the average household income, the average age a household head, and the average household size in the market. Unobserved by the analyst, the market/choice set/time specific quality of brand $y$, which is potentially correlated with $\rand{p}_{y,j}$, is captured by $\xi_{y,D}+\Delta\rands{\xi}_{y,D,j,t}$. The first term is the mean quality of a product $y$ in a choice set $D$. The second term is the mean-zero choice set/time/brand specific deviation from that quality; $\rands{\varepsilon}_{y,D,t}$ is the additive random shock that is independent from all other variables. These shocks are i.i.d. with a Type I extreme-value distribution. These assumptions reduce the model to the well-known (multinomial) Logit model. 
\par
Note that in this setup, Assumption~\ref{ass: CIP} is satisfied if $\Delta\rands{\xi}_{y,D,j,t}$s are conditionally independent across time conditionally on choice sets, the observables, and the market identity. Again, given the high-frequency of our dataset, we think the conditional independence assumption is reasonable in our setting as roughly $3$ weeks of purchases would not be enough time to affect habit formation, which is the main source of inertia.
\par
Here, for simplicity, we assume the Logit model and that $\alpha_{y,D}$, $\gamma_{y,D}$, and $\beta_{D}$ are fixed parameters. The model can be extended to the Generalized Extreme Value model \citep{mcfadden1978modelling}, which includes the Nested Logit model, and to the case when coefficients are random (e.g, \citealp{nevo2001measuring}). Moreover, since we are estimating shares conditional on the market, after the shares are estimated, one can add as many market level covariates to the model as we want.
\par
The parameter of interest is $\beta_{D}$ that can vary with choice sets. This parameter captures the price elasticity of demand for all brands. Note that since it is indexed by the choice set $D$ we allow for correlation between preferences and choice sets. We use variation across time and markets to estimate $\beta_{D}$.\footnote{One can allow $\beta$ to vary across time. In this case, one would have to use only the market variation.}
\par
Our parametric specification implies that for any $y,\bar{y}\in D$
\[
\Delta\rands{\xi}^*_{y,\bar{y},D,j,t}=\Delta\rands{\xi}_{y,D,j,t}-\Delta\rands{\xi}_{\bar{y},D,j,t}=\log\left(\dfrac{\rand{s}_{y,D,j,t}}{\rand{s}_{\bar{y},D,j,t}}\right)-\alpha^*_{y,\bar{y},D}-\beta_D(\rand{p}_{y,j}-\rand{p}_{\bar{y},j})-\rand{r}_{j}\tr\gamma^*_{y,\bar{y},D},
\]
where $\alpha^*_{y,\bar{y},D}=\alpha_{y,D}-\alpha_{\bar{y},D}+\xi_{y,D}-\xi_{\bar{y},D}$ and $\gamma^*_{y,\bar{y},D}=\gamma_{y,D}-\gamma_{\bar{y},D}$. Thus, if $\Delta\rands{\xi}^*_{y,\bar{y},D,j,t}$ was not correlated with prices, then we could have used the ordinary least squares estimator to consistently estimate $\beta_D$. However, because of the price endogeneity, we use instruments and the two-step efficient Generalized Method of Moments (GMM) estimator. In particular, following \citet{berry1995automobile,nevo2000practitioner}, and \citet{nevo2001measuring}, we construct two instruments: average product characteristics (i.e., size) of competing brands and average across neighboring markets price of the brand.\footnote{The details of construction of instruments can be found in our replication files.}
\par
First, we estimate the model assuming that there is no variation in choice sets. That is, every DM faces all $5$ brands (Naive estimator). Next we estimate $\beta_D$ using our procedure for two most popular choice sets: \{CTL, GM, K, O, Q\} and \{CTL, GM, K, O\}. In both cases, we use other brands of cereal (O) as the base option $\bar{y}$. The results of estimation are presented in Table~\ref{table: estimates of beta}.\footnote{In Appendix~\ref{app: empirical application}, we construct the standard errors for these estimates under the assumption that there is no estimation error in the estimated market shares and all uncertainty is coming from variation in observed covariates.}

\begin{table}[h]
\centering
\begin{threeparttable}
\centering
\caption{Estimates of $\beta_D$}\label{table: estimates of beta}
\begin{tabular}{lccc}
\hline
\hline
   &Naive & \{CTL, GM, K, O, Q\} & \{CTL, GM, K, O\}\\ 
$\beta_D$ & -16.11 & -5.84 & -56.46\\
\hline
\end{tabular}
\vspace{1ex}
 \begin{tablenotes}
 \item {\footnotesize Notes: Results are rounded to 2 digits.}
 \end{tablenotes}
\end{threeparttable}
\end{table}
There is substantial heterogeneity between 2 types of DMs. The ones that consider all 5 brands have the price coefficient approximately 3 times smaller in magnitude than the naive estimate. The DMs who consider all brands but Quaker have substantially larger in magnitude price coefficient. Given these estimates of the price coefficient, we can compute the implied own-price elasticities under the assumption that the distribution over the choice sets $m$ does not depend on prices.\footnote{See \citet{goeree2008limited} for similar exclusion restrictions.} Formally, the own-price elasticity of brand $y$ in market $j$ at time $t$ is defined as\footnote{Without exclusion restrictions, one would also need to take into account the derivatives of $m$ with respect to prices.}
\[
\mathrm{Elas}_{y,j,t}=\dfrac{p_{y,j}}{s_{y,j,t}}\dfrac{\partial s_{y,j,t}}{\partial{p_{y,j}}}=\dfrac{p_{y,j}}{s_{y,j,t}}\sum_{D}\dfrac{\partial s_{y,D,j,t}}{\partial{p_{y,j}}}m(D|j),
\]
where $s_{y,j,t}$ is the observed share of brand $y$ in market $j$ at time $t$. If there is no choice set variation, then, under our parametrization,
\[
\mathrm{Elas}_{y,j,t}=\beta_{\mathrm{Naive}}p_{y,j}(1-s_{y,j,t}).
\]
With choice set variation and under the assumption that prices do not affect choice set probabilities,
\[
\mathrm{Elas}_{y,j,t}=\sum_{D}\dfrac{s_{y,D,j,t}}{s_{y,j,t}}\mathrm{Elas}_{y,D,j,t}m(D|j),
\]
where
\[
\mathrm{Elas}_{y,D,j,t}=\beta_{D}p_{y,j}(1-s_{y,D,j,t}).
\]

Since elasticities may not be constant across markets we report own-price elasticities for the largest in terms of observations market (Market 1) in Table~\ref{table: elast}. In the first column, we use estimates of the price coefficient assuming that there is no choice set variation (Naive). The second column is computed using our estimates of the price coefficients for different choice sets. The third and the last column report elasticities for those who consider all 5 brands or do not consider Quaker, respectively (i.e., $\mathrm{Elas}_{y,D,j,t}$).
\begin{table}[h]
\centering
\begin{threeparttable}
\centering
\caption{Estimates of Own-Price Elasticities in Market 1}\label{table: elast}
\begin{tabular}{lcccc}
\hline
\hline
   &Naive & Choice Set Variation& \{CTL, GM, K, O, Q\} & \{CTL, GM, K, O\}\\ 
CTL & -1.98 & -0.98 & -0.76 & -7.51\\
GM  & -2.34 & -1.39 & -0.83 & -10.06\\
K   & -2.03 & -1.08 & -0.67 & -8.81\\
O   & -2.25 & -2.35 & -0.87 & -3.39\\
Q   & -2.4  & -0.7  & -0.85 & 0\\
\hline
\end{tabular}
\vspace{1ex}
\begin{tablenotes}
\item {\footnotesize Notes: The first column is computed assuming that all consumers face all 5 brands. The second column is computed assuming choice set variation. The third column is computed for those consumers who consider all 5 brands. The last column is computed for those consumers who do not consider Quaker. Results are rounded to 2 digits.}
\end{tablenotes}
\end{threeparttable}
\end{table}
\par
The estimates of the own-price elasticities that assumes no choice set variation are similar to ones in \citet{nevo2001measuring}.\footnote{In Appendix~\ref{app: empirical application}, we report the median across markets own-price elasticities. The results are qualitatively the same.} However, the demand of those considering all brands is substantially less elastic than those who do not consider Quaker. That is, we find substantial unobserved heterogeneity in how consumers react to price changes. As a result of this heterogeneity, the implied own-price elasticity that takes into account the choice set variation is smaller for almost all brands. For Quaker, given that it is not considered by a large group of consumers (about 83 percent), the difference is more than three fold. 
\par
Estimating own-price elasticities without considering hidden categorization/menu variation could lead to biased estimates. Here, frequent buyers purchase a cereal from a particular category of brands. For instance, some consumers could always avoid Quaker cereal or only consider General Mills. Others, however, may consider everything. In general, these frequent buyers of RTE cereal may have strong opinions about what they like to consider and what they avoid. Also, they are exposed to advertisement, promotions, and other factors that affect the category of brands they consider. Our approach remains completely flexible with respect to the particular story that leads to the formation of a category of brands but imposes a sparsity restriction. Namely, conditional on a market, there can be at most $5$ choice sets in each market.

\section{Conclusion}\label{sec:conclusion}
In this paper, we show that observing three or more choices from the same latent choice set is sufficient to nonparametrically identify and consistently estimate the joint distribution of choice sets and choices in discrete-choice models when choice sets are not observable. Our main result requires a linear independence condition on the conditional distribution of choices. This condition is satisfied when either the panel is long enough or the support of choice sets is sparse. The application of our computationally efficient estimator to a scanner dataset indicates that there is a substantial unobserved choice set heterogeneity and correlation between preferences and choice sets that can bias estimates of the standard parameters of interest (e.g., own-price elasticities).

\phantomsection\addcontentsline{toc}{section}{\refname}\bibliography{considerationsetsfield}
\appendix

\section{Proofs}\label{app:proofs}
\subsection{Proof of Proposition~\ref{prop: suff cond for linindep}}
Fix some $x\in X$ and $\mathcal{T}^*$. To simplify the exposition, we drop $x$ and $\mathcal{T}^*$ from the notation. Note that if the linear independence condition is satisfied when $\mathcal{D}_x=2^{\mathcal{Y}}\setminus\emptyset$, then it is automatically satisfied for any smaller $\mathcal{D}_x$. Hence, without loss of generality, we assume that $\mathcal{D}_x=2^{\mathcal{Y}}\setminus\emptyset$. 
\par
Consider the following order of elements in $\mathcal{D}_x$. First, we have all singleton sets in an arbitrary order, next we have all sets of cardinality $2$ in arbitrary order, so on and so forth. For any two sets $D_k$ and $D_m$ such that $k<m$, $D_k$ is either a strict subset of $D_m$ or has no element in common with it. 
\par
Given the above order over sets $D$, consider the following sequence of $2^Y-1$ vectors in $\mathcal{Y}^K$. For any $D_k$ take $y^K_{k}$ such that every element in $D_k$ is some component of $y_k^K$. Since $K\geq Y$ and $\abs{D_k}\leq Y$ such $y^K$ always exists. Moreover, for any $D_k\neq D_l$ it has to be true that $y_k^K\neq y^K_{l}$.
\par
Consider the matrix $\mathcal{G}$ of size $(2^Y-1)\times(2^Y-1)$ such that $(j,k)$-element of it is
\[
\mathcal{G}_{j,k}=G(y^K_{j}\mid D_k).
\]
Note that this matrix is upper-triangular. Indeed, take any $j,k$ such that $j>k$. Since $j>k$, then $D_k$ is either a strict subset or does not overlap with $D_j$. Hence, there exists a component of $y_j^K$ that is not an element of $D_k$. This means that the probability of observing a sequence $y_j^K$ given set $D_k$ is zero. That is, $\mathcal{G}_{j,k}=0$ if $j>k$. Assumption~\ref{ass:regularity} implies that the diagonal elements, $\mathcal{G}_{j,j}$, are nonzero since any element of $D_j$ can be observed with positive probability. Since $\mathcal{G}$ is upper-triangular with nonzero diagonal elements, it is of full column rank (the determinant of $\mathcal{G}$ equals to the product of the diagonal elements). 
\par
Adding more rows to $\mathcal{G}$ does not change its column rank. Hence, the linear independence condition is satisfied. The fact that the choice of $x$ and $\mathcal{T}^*$ was arbitrary completes the proof.

\subsection{Proof of Proposition~\ref{prop: suff cond for linindep2}}
\emph{(i).} Fix some $x\in X$ and $\mathcal{T}^*$. To simplify the exposition, we drop $x$ and $\mathcal{T}^*$ from the notation. Note that if the linear independence condition is satisfied when $\mathcal{D}_x=\{\{y_1\}, \{y_1,y_2\}, \{y_1,y_2,y_3\},\dots,\mathcal{Y}\}$, then it is automatically satisfied for any smaller $\mathcal{D}_x$. So, without loss of generality, we assume that $\abs{\mathcal{D}_x}=Y$. 
\par
Nestedness implies that $D_{k}\subseteq D_{k+1}$ for all $k$. Let $\{y^K_j\}_{j=1}^Y$ be a sequence in $\mathcal{Y}^K$ such that $y^K_j=(j,j,\dots,j)\tr$. Recall that $D_k=\{y_1,y_2,\dots,y_k\}$. Consider the matrix $\mathcal{G}$ of size $Y\times Y$ such that $(j,k)$-element of it is
\[
\mathcal{G}_{j,k}=G(y^K_{j}\mid D_k).
\]
Note that this matrix is upper-triangular. Indeed, take any $j,k$ such that $j>k$. Hence, none of the components of $y_j^K$ are elements of $D_k$. This means that the probability of observing a sequence $y_j^K$ given set $D_k$ is zero. That is, $\mathcal{G}_{j,k}=0$ if $j>k$. Assumption~\ref{ass:regularity} implies that the diagonal elements, $\mathcal{G}_{j,j}$, are nonzero since any element of $D_j$ can be observed with positive probability. Since $\mathcal{G}$ is upper-triangular with nonzero diagonal elements, it is of full column rank (the determinant of $\mathcal{G}$ equals to the product of the diagonal elements). Adding more rows to $\mathcal{G}$ does not change its column rank. Hence, the linear independence condition is satisfied. The fact that the choice of $x$ and $\mathcal{T}^*$ was arbitrary completes the proof.
\par
\emph{(ii).} Fix some $x\in X$ and $\mathcal{T}^*$. To simplify the exposition, we drop $x$ and $\mathcal{T}^*$ from the notation. Let $\{y^K_k\}_{k=1}^{\abs{\mathcal{D}_x}}$ be a sequence in $\mathcal{Y}^K$ such that $y^K_k=(y_k,y_k,\dots,y_k)\tr$, where $y_k\in \mathcal{Y}$ are from condition (ii) of the proposition. Consider the matrix $\mathcal{G}$ of size $\abs{\mathcal{D}_x}\times\abs{\mathcal{D}_x}$ such that $(j,k)$-element of it is
\[
\mathcal{G}_{j,k}=G(y^K_{j}\mid D_k).
\]
Note that this matrix is diagonal. Indeed, take any $j,k$ such that $j\neq k$. Hence, none of the components of $y_j^K$ are elements of $D_k$. This means that the probability of observing a sequence $y_j^K$ given set $D_k$ is zero. That is, $\mathcal{G}_{j,k}=0$ if $j\neq k$. Assumption~\ref{ass:regularity} implies that the diagonal elements, $\mathcal{G}_{j,j}$, are nonzero since any element of $D_j$ can be observed with positive probability. Since $\mathcal{G}$ is diagonal with nonzero diagonal elements, it is of full column rank (the determinant of $\mathcal{G}$ equals to the product of the diagonal elements). Adding more rows to $\mathcal{G}$ does not change its column rank. Hence, the linear independence condition is satisfied. The fact that the choice of $x$ and $\mathcal{T}^*$ was arbitrary completes the proof.

\subsection{Proof of Theorem~\ref{thm:identification}}
Fix some $x\in\X$ and $t\in\mathcal{T}$. Take two disjoint subsets of size $K$, $\mathcal{T}^*$ and $\mathcal{T}^{**}$, that do not contain $t$. Since $K$ is the biggest integer that is less or equal to $(T-1)/2$ such sets always exist. Let $g:\mathcal{Y}^K\to \bar{\mathcal{Y}}=\{1,2,\dots,Y^K\}$ be any one-to-one mapping. Define two random variable on $\bar{\mathcal{Y}}$: $\rand{z}_1=g(\rand{y}(\mathcal{T}^{*}))$ and $\rand{z}_2=g(\rand{y}(\mathcal{T}^{*}))$. To simplify the exposition, we drop $x$ and $\mathcal{T}^*$ from the notation. Define the following matrices
\begin{align*}
    L_{1,2}&=\left[\Prob{\rand{z}_1=i,\rand{z}_2=j}\right]_{i,j\in\bar{\mathcal{Y}}},\\
    L_{1|\D}&=\left[\Prob{\rand{z}_1=i\mid \rand{\D}=\D_k}\right]_{i\in\bar{\mathcal{Y}},k=1,\dots,d_\D},\\
    L_{2|\D}&=\left[\Prob{\rand{z}_2=i\mid \rand{\D}=\D_k}\right]_{i\in\bar{\mathcal{Y}},k=1,\dots,d_\D},\\
    A_{\D}&=\mathrm{diag}\left((\Prob{\rand{\D}=\D_k})_{k=1,\dots,d_\D}\right)=\mathrm{diag}\left((m(D_k))_{k=1,\dots,d_\D}\right),
\end{align*}
where $\mathrm{diag}(z)$ is a diagonal matrix with vector $z$ on the diagonal.
\par
\noindent\emph{Step 1.} In this step, we show how to identify the number of choice sets that are considered with positive probability. By the law of total probability, under Assumption~\ref{ass: CIP},
\begin{align*}
\Prob{\rand{z}_1=i,\rand{z}_2=j}&=\sum_k\Prob{\rand{z}_1=i,\rand{z}_2=j\mid \rand{\D}=\D_k}\Prob{\rand{\D}=\D_k}\\
&=\sum_k\Prob{\rand{z}_1=i\mid \rand{\D}=\D_k}\Prob{\rand{z}_2=j\mid \rand{\D}=\D_k}\Prob{\rand{\D}=\D_k}.
\end{align*}
Or in matrix notation
\[
L_{1,2}=L_{1|\D}A_\D L_{2|\D}\tr.
\]
Under Assumption~\ref{ass:linearindependence} the maximal number of the points in the support of $\rand{\D}$ is equal to the number of the possible outcomes. That is, $d_\D\leq\abs{\bar{\mathcal{Y}}}$. 
\par
Next, note that Assumption~\ref{ass:linearindependence} implies that $L_{1|\D}$ and $L_{2|\D}$ have full column rank ($d_\D$). Hence, using the properties of the rank operator we can conclude that
\begin{align*}
    \mathrm{rank}\left(L_{1,2}\right)=\mathrm{rank}\left(L_{1|\D}A_\D L_{2|\D}\tr\right)=\mathrm{rank}\left(A_\D L_{2|\D}\tr\right)=\mathrm{rank}\left(A_\D\right)=d_\D.
\end{align*}
That is, the rank of $L_{1,2}$ is equal to $d_D=\abs{\mathcal{D}_x}$. Hence, since $L_{1,2}$ is observed (can be consistently estimated), we can identify (consistently estimate) the number of choice sets that DMs are using.
\par
\noindent\emph{Step 2.} Knowing $d_D$ and the fact that $L_{1|D}$ and $L_{2|D}$ have full column rank, we take a collection of alternatives in $\bar{\mathcal{Y}}$, $\{\tilde z_k\}_{k=1}^{d_D}$, such that the following observable modification of $L_{1,2}$ is nonsingular (have full rank):
\begin{align*}
    \tilde L_{1,2}&=\left[\Prob{\rand{z}_1= \tilde z_i,\rand{z}_2=\tilde z_j}\right]_{i,j\in\{1,\dots,d_D\}}.
\end{align*}
Such collection $\{\tilde z_k\}_{k=1}^{d_D}$ always exists since one can always find $d_D$ linearly independent rows of $L_{1|D}$. Indeed, similar to Step 1
\[
\tilde L_{1,2}=\tilde L_{1|D} A_D\tilde L_{2|D}\tr,
\]
where 
\begin{align*}
    \tilde L_{1|\D}&=\left[\Prob{\rand{z}_1=\tilde z_i|\rand{\D}=\D_k}\right]_{i,k\in\{1,\dots,d_D\}},\\
    \tilde L_{2|\D}&=\left[\Prob{\rand{z}_2=\tilde z_i|\rand{\D}=\D_k}\right]_{i,k\in\{1,\dots,d_D\}}.
\end{align*}
Since $\tilde L_{1|\D}$ and $\tilde L_{2|\D}$ are nonsingular, it implies that $\tilde L_{1,2}$ is nonsingular as well ($A_D$ has rank $d_D$). 

\noindent\emph{Step 3}. This step is based on \citet{hu2008identification} and \citet{hu2013identification}. Fix some $y\in\mathcal{Y}$ and define
\begin{align*}
    \tilde L_{1,D}&=\left[\Prob{\rand{z}_1=\tilde z_i,\rand{\D}=\D_k}\right]_{i,k\in\{1,\dots,d_D\}},\\
    \tilde L_{2,1,y}&=\left[\Prob{\rand{z}_2=\tilde z_i,\rand{z}_1=\tilde z_j,\rand{y}_t=y}\right]_{i,j\in\{1,\dots,d_D\}},\\
    A_{y|\D}&=\mathrm{diag}\left((\Prob{\rand{y}_t=y|\rand{\D}=\D_k})_{k\in\{1,\dots,d_D\}}\right)=\mathrm{diag}\left((F_3^{\rumo}(\y|\D_k))_{k\in\{1,\dots,d_D\}}\right).
\end{align*}
By the law of total probability, under Assumption~\ref{ass: CIP},
\begin{align*}
\Prob{\rand{z}_1=\tilde z_i,\rand{z}_2=\tilde z_j}&=\sum_k\Prob{\rand{z}_1=\tilde z_i,\rand{z}_2=\tilde z_j|\rand{\D}=\D_k}\Prob{\rand{\D}=\D_k}\\
&=\sum_k\Prob{\rand{z}_2= \tilde z_j|\rand{\D}=\D_k}\Prob{\rand{z}_1=\tilde z_i,\rand{\D}=\D_k}.
\end{align*}
Hence, in matrix notation we get
\[
\tilde L_{1,2}\tr=\tilde L_{2|\D}\tilde L_{1,D}\tr.
\]
Since, by construction in Step 2, $\tilde L_{2|\D}$ is nonsingular, we have that
\begin{align}\label{eq:1}
\tilde L_{1,D}\tr= \tilde L_{2|\D}^{-1}\tilde L_{1,2}\tr.
\end{align}
\par
Similarly to the previous calculations, under Assumption~\ref{ass: CIP}, we have
\[
\tilde L_{2,1,y}=\tilde L_{2|\D}A_{y|\D}\tilde L_{1,D}\tr.
\]
Combining the latter with equation (\ref{eq:1}) we get the following eigenvector-eigenvalue decomposition of $R_y=\tilde L_{2,1,y}\left(\tilde L_{1,2}\tr\right)^{-1}$
\begin{equation}\label{eq:decomposition}
R_y=\tilde L_{2|\D}A_{y|\D}\tilde L_{2|\D}^{-1}.
\end{equation}
\par
\noindent\emph{Step 4}. Note that in the decomposition (\ref{eq:decomposition}) the change in $y$ does not affect eigenvectors of $R_y$, but affects its eigenvalues. For $R_y$ let $\{(\eta_k,\lambda_{y,k})\}_{k=1}^{d_D}$ denote the set of its eigenvectors and eigenvalues. To pin down eigenvectors uniquely note that it suffices to pick those that belong to a simplex (each one of them should sum up to 1). In contrast to the existing results (e.g. \citealp{hu2013identification}), we do not use these eigenvectors to identify $L_{2|D}$ since $\tilde L_{2|D}$ is only a submatrix of $L_{2|D}$.
\par
Take $y=1$ and fix the set of eigenvectors of $R_{1}$, $\{\eta_k\}_{k=1}^{d_D}$. Stack them in any order to a matrix $\Lambda_1$. Then we can compute 
\[
A^*_{y|D}=\Lambda_1^{-1}R_y\Lambda_1
\]
for every $y$. Since the order of eigenvalues is fixed, the diagonal entries of $A^*_{y|D}$ correspond to the same sets. Note that $y\in D$ if and only if $F_t^{\rumo}(y|D)>0$. Thus, we can identify the identity of choice sets and $F_t^{\rumo}(y|D)$ for every $y$ and $D$.
Since the choice of $t$ was arbitrary, we can identify $F_t^{\rumo}$ for all $t\in\mathcal{T}$. Hence, we can identify the conditional distributions of $\rand{z}_1$ and $\rand{z}_2$ conditional on $\rand{D}$. Thus, we identify $L_{2|D}$.
\par
\noindent\emph{Step 5}. 
Finally, let $m=(m(D))_{D\in\mathcal{D}_x}$, then
\[
m=L_{1,D}\tr\cdot\iota,
\]
where $\iota$ is the vector of ones. Hence,
\[
L_{1,2}\tr\iota=L_{2|D}L_{1,D}\tr\cdot\iota= L_{2|D}m.
\]
Since $L_{1,2}$ is observed (can be consistently estimated), and $L_{2|D}$ is constructively identified and has full column rank, we also identify the distribution of choice sets. The fact that the choice of $x$ was arbitrary completes the proof.

\subsection{Proof of Lemma~\ref{lem: dynam cond indep}}
Assume that $K_d=1$. Also, to simplify the exposition, we drop $\rand{D}$ and $\rand{x}$ and write $\Prob{y_t}$ instead of $\Prob{\rand{y}_t=y_t}$. First, note that after applying several times Bayes' theorem and Assumption~\ref{ass: markov}, we get that
\begin{align*}
\Prob{y_1\mid y_4,y_3,y_2}&=\dfrac{\Prob{y_4,y_3,y_2,y_1}}{\Prob{y_4,y_3,y_2}}=\dfrac{\Prob{y_4\mid y_3,y_2,y_1}\Prob{y_3\mid y_2,y_1}\Prob{y_2,y_1}}{\Prob{y_4\mid y_3,y_2}\Prob{y_3\mid y_2}\Prob{y_2}}\\
&=\dfrac{\Prob{y_4\mid y_3}\Prob{y_3\mid y_2}\Prob{y_2,y_1}}{\Prob{y_4\mid y_3}\Prob{y_3\mid y_2}\Prob{y_2}}=\dfrac{\Prob{y_2,y_1}}{\Prob{y_2}}=\Prob{y_1\mid y_2}.
\end{align*}
Similarly, $\Prob{y_1\mid y_4,y_2}=\Prob{y_1\mid y_2}$. Hence,
\begin{align*}
\Prob{y_3\mid y_4,y_2,y_1}&=\dfrac{\Prob{y_3,y_1\mid y_4,y_2}}{\Prob{y_1\mid y_4,y_2}}=\dfrac{\Prob{y_1\mid y_4,y_3,y_2}\Prob{y_3\mid y_4,y_2}}{\Prob{y_1\mid y_2}}\\
&=\dfrac{\Prob{y_1\mid y_2}\Prob{y_3\mid y_4,y_2}}{\Prob{y_1\mid y_2}}=\Prob{y_3\mid y_4,y_2}.
\end{align*}
As a result,
\begin{align*}
\Prob{y_5,y_3,y_1\mid y_4,y_2}&=\Prob{y_5\mid y_4,y_3,y_2,y_1}\Prob{y_3\mid y_4,y_2,y_1}\Prob{y_1\mid y_4,y_2}\\
&=\Prob{y_5\mid y_4,y_2}\Prob{y_3\mid y_4,y_2}\Prob{y_1\mid y_4,y_2}.
\end{align*}
Thus, $\rand{y}_5$, $\rand{y}_3$, and $\rand{y}_1$ are conditionally independent conditional on $\rand{y}_4$ and $\rand{y}_2$. For $K_d>1$ one just need to relabel $y_5$ as $y_{2K_d+3}$, $y_4$ as $y_{2K_d+2}$, $y_{3}$ as $y(\mathcal{T}_2)$, $y_2$ as $y_{K_d+1}$, and $y_{1}$ as $y(\mathcal{T}_1)$, and apply the above arguments.

\subsection{Proof of Proposition~\ref{prop: dynamic identification}}
Similar to the proof of Theorem~\ref{thm:identification} let $\rand{z}_i=g(\mathcal{T}_{d,i})$, $i=1,2$, and $t=2K_d+3$. Since conditional on $\rand{y}_{K_d+1}$ and $\rand{y}_{2K_d+2}$, $\rand{z}_1$, $\rand{z}_2$, and $\rand{y}_{t}$ are independent, we can repeat the steps of the proof of Theorem~\ref{thm:identification} and identify $\Prob{\rand{y}_{2K_d+3}=y\mid \rand{y}_{2K_d+2}=y,\rand{D}=D,\rand{x}=x}$ for all $y,y',D$, and $x$. Thus, under Assumption~\ref{ass: dist stab}, we can identify $F_d^{\rumo}$. Knowing $F_d^{\rumo}$, following the steps of the proof of Theorem~\ref{thm:identification}, we can also identify $m$.

\section{Monte Carlo Simulations}\label{app: MC simulations}
The aim of this section is to analyse the finite sample performance of the estimator we propose in Section~\ref{sec: estimation}. First, we define the data generating processes (DGPs) used in simulations. In all experiments we assume that there are no covariates, $T=3$, $Y=5$, $d_{\mathcal{D}}=5$, and $F^{\rumo}_1=F^{\rumo}_2=F^{\rumo}_3$. Every DGP is characterized by two matrices: $Pyd\in\Real^{5\times5}$ and $Pd\in\Real^5$.  $Pyd$ and $Pd$ is such that $Pyd_{y,j}=\Prob{\rand{y}_t=y\mid D_j}$ and $Pd_j=\Prob{\rand{D}=D_j}$.
\par
\noindent\textbf{DGP1:} 
\[
Pyd=\left(\begin{array}{ccccc}
    1& 0.6& 0.5& 0.4& 0.2\\
    0& 0.4& 0& 0& 0\\
    0& 0& 0.5& 0& 0\\
    0& 0& 0& 0.6& 0\\
    0& 0& 0& 0& 0.8
\end{array}\right),\quad 
Pd=\left(\begin{array}{c}
    0.2\\
    0.15\\
    0.3\\
    0.15\\
    0.2
\end{array}\right)
\]
\noindent\textbf{DGP2:} 
\[
Pyd=\left(\begin{array}{ccccc}
    1& 0.6& 0.5& 0.25& 0.1\\
    0& 0.4& 0.2& 0.35& 0.25\\
    0& 0& 0.3& 0.25& 0.15\\
    0& 0& 0& 0.15& 0.3\\
    0& 0& 0& 0& 0.2
\end{array}\right),\quad 
Pd=\left(\begin{array}{c}
    0.2\\
    0.15\\
    0.3\\
    0.15\\
    0.2
\end{array}\right)
\]
The results for estimation of choice sets is presented in Section~\ref{sec:simulations}. Tables~\ref{table: bias,M,dgp1}-~\ref{table: rmse,F,dgp2} present the bias and the mean-squared-error of estimation of $m$ and $F_1^{\rumo}$. Every experiment was conducted $1000$ times. Since some elements of matrix $Pyd$ are zeros and elements of every column sum up to one, for estimates of $F_1^{\rumo}$ we only report the nonzero, linearly independent elements. For instance, for DGP1 we only report estimates of $(1,2)$, $(1,3)$, $(1,4)$, and $(1,5)$ elements of $Pyd$. The estimators perform well even in the samples of a moderate size. As expected, both the bias and the root-mean-squared-error decrease with the sample size.

\begin{table}[h]
\centering
\begin{threeparttable}
\centering
\caption{Bias in Estimating $m$. DGP1 ($\times 10^{-5}$)}\label{table: bias,M,dgp1}
\begin{tabular}{cccccc}
\hline
\hline
Sample Size$\setminus$Set    & $D_1$ & $D_2$ & $D_3$ & $D_4$ & $D_5$\\ 
\hline
2000  &  -112.6 & 35.1 & 43.8 & 35.7 & -2\\
5000  & -30.4 & 18.7 & 8.7 & 1.8 & 1.2 \\
10000 & -2.2 & -8.1 & 8.3 & -1.3 & 3.3\\
50000 & 4.8 & -12.5 & -0.004 & 6.2 & 1.6\\
\hline
\end{tabular}
\vspace{1ex}
\begin{tablenotes}
\item {\footnotesize Notes: Number of replications=1000. Results are rounded to 6 digits.}
\end{tablenotes}
\end{threeparttable}
\end{table}

\begin{table}[h]
\centering
\begin{threeparttable}
\centering
\caption{Root Mean Squared Error in Estimating $m$. DGP1 ($\times 10^{-3}$)}\label{table: rmse,M,dgp1}
\begin{tabular}{cccccc}
\hline
\hline
Sample Size$\setminus$Set    & $D_1$ & $D_2$ & $D_3$ & $D_4$ & $D_5$\\ 
\hline
2000  & 13.1 & 10.5 & 12.5 & 8.5 & 9.2\\
5000  & 8    & 6.6  & 7.7  & 5.4 & 5.9\\
10000 & 5.8  & 4.8  & 5.2  & 3.7 & 4.2\\
50000 & 2.7  & 2.1  & 2.4  & 1.7 & 1.9\\
\hline
\end{tabular}
\vspace{1ex}
\begin{tablenotes}
\item {\footnotesize Notes: Number of replications=1000. Results are rounded to 4 digits.}
\end{tablenotes}
\end{threeparttable}
\end{table}

\begin{table}[h]
\centering
\begin{threeparttable}
\centering
\caption{Bias in Estimating $m$. DGP2 ($\times 10^{-3}$)}\label{table: bias,M,dgp2}
\begin{tabular}{cccccc}
\hline
\hline
Sample Size$\setminus$Set    & $D_1$ & $D_2$ & $D_3$ & $D_4$ & $D_5$\\ 
\hline
2000  & -16.7 & -9.7 & -54.2 & 37.7 & 42.9\\
5000  & -11.8 & -7.4 & -65.8 & 53.4 & 31.6\\
10000 & -6.7 & -2.3 & -50.1 & 40.7 & 18.4\\
50000 & -0.6 & 0.4 & -6.5 & 4.3 & 2.4\\
\hline
\end{tabular}
\vspace{1ex}
\begin{tablenotes}
\item {\footnotesize Notes: Number of replications=1000. Results are rounded to 4 digits.}
\end{tablenotes}
\end{threeparttable}
\end{table}

\begin{table}[h]
\centering
\begin{threeparttable}
\centering
\caption{Root Mean Squared Error in Estimating $m$. DGP2 ($\times 10^{-2}$)}\label{table: rmse,M,dgp2}
\begin{tabular}{cccccc}
\hline
\hline
Sample Size$\setminus$Set    & $D_1$ & $D_2$ & $D_3$ & $D_4$ & $D_5$\\ 
\hline
2000  & 5.3 & 4.7 & 10.4 & 10.7 & 5.1\\
5000  & 4.0 & 3.3 & 10.6 & 10.1 & 4.0\\
10000 & 2.6 & 1.8 &  8.3 &  7.5 & 2.6\\
50000 & 0.3 & 0.5 &  1.5 &  1.2 & 0.6\\
\hline
\end{tabular}
\vspace{1ex}
\begin{tablenotes}
\item {\footnotesize Notes: Number of replications=1000. Results are rounded to 3 digits.}
\end{tablenotes}
\end{threeparttable}
\end{table}

\begin{table}[h]
\centering
\begin{threeparttable}
\centering
\caption{Bias in Estimating $F^{\rumo}_1$. DGP1 ($\times 10^{-4}$)}\label{table: bias,F,dgp1}
\begin{tabular}{ccccccccc}
\hline
\hline
Sample Size$\setminus (i,j)$    & $1,2$ & $1,3$ & $1,4$ & $1,5$ \\ 
\hline
2000  & 0.9 &  2.7 & -0.4 & -1.8\\
5000  & 3.7 & -4.4 &  8.5 &  1.3\\
10000 & 0.9 & -1.6 &  2.9 & -0.6\\
50000 &-2.3 & -2.1 & -0.0 & -0.3\\
\hline
\end{tabular}
\vspace{1ex}
\begin{tablenotes}
\item {\footnotesize Notes: Number of replications=1000. Results are rounded to 5 digits. Only estimated elements of $Pyd$ are displayed. Only linearly independent estimated elements of $\hat{F}^{\rumo}_1$ are displayed.}
\end{tablenotes}
\end{threeparttable}
\end{table}

\begin{table}[h]
\centering
\begin{threeparttable}
\centering
\caption{Root Mean Squared Error in Estimating $F^{\rumo}_1$. DGP1 ($\times 10^{-2}$)}\label{table: rmse,F,dgp1}
\begin{tabular}{ccccccccc}
\hline
\hline
Sample Size$\setminus (i,j)$    & $1,2$ & $1,3$ & $1,4$ & $1,5$ \\ 
\hline
2000  & 3.3 & 2.3 & 3.0 & 2.3\\
5000  & 2.1 & 1.4 & 2.0 & 1.5\\
10000 & 1.5 & 1.0 & 1.4 & 1.0\\
50000 & 0.7 & 0.4 & 0.6 & 0.4\\
 \hline
\end{tabular}
\vspace{1ex}
\begin{tablenotes}
\item {\footnotesize Notes: Number of replications=1000. Results are rounded to 3 digits. Only linearly independent estimated elements of $\hat{F}^{\rumo}_1$ are displayed.}
\end{tablenotes}
\end{threeparttable}
\end{table}

\begin{table}[h]
\centering
\begin{threeparttable}
\centering
\caption{Bias in Estimating $F^{\rumo}_1$. DGP2 ($\times 10^{-2}$)}\label{table: bias,F,dgp2}
\begin{tabular}{ccccccccccccccc}
\hline
\hline
Sample Size$\setminus (i,j)$    & $1,2$ & $1,3$ & $2,3$ & $1,4$ & $2,4$ & $3,4$ & $1,5$ & $1,5$ & $3,5$ & $4,5$\\ 
\hline
2000  & 0.9 & 3.3 & -2.3 & 7.7 & -0.7 & -0.1 & 1.4 & 1.6 & 1.3 & -0.7\\
5000  & 1.9 & 3.7 & -2.9 & 8.1 & -0.6 & -0.4 & 0.9 & 1.2 & 1.0 & -0.2\\
10000 & 1.2 & 3.3 & -2.7 & 5.4 & -0.4 &  0.0 & 0.4 & 0.7 & 0.7 & -0.0\\
50000 & 0.2 & 0.3 & -0.4 & 0.6 &  0.2 &  0.2 & 0.1 & 0.1 & 0.0 &  0.1\\
\hline
\end{tabular}
\vspace{1ex}
\begin{tablenotes}
\item {\footnotesize Notes: Number of replications=1000. Results are rounded to 3 digits. Only estimated elements of $Pyd$ are displayed. Only linearly independent estimated elements of $\hat{F}^{\rumo}_1$ are displayed.}
\end{tablenotes}
\end{threeparttable}
\end{table}

\begin{table}[h]
\centering
\begin{threeparttable}
\centering
\caption{Root Mean Squared Error in Estimating $F^{\rumo}_1$. DGP2 ($\times 10^{-2}$)}\label{table: rmse,F,dgp2}
\begin{tabular}{ccccccccccccccc}
\hline
\hline
Sample Size$\setminus (i,j)$    & $1,2$ & $1,3$ & $2,3$ & $1,4$ & $2,4$ & $3,4$ & $1,5$ & $1,5$ & $3,5$ & $4,5$\\ 
\hline
2000  & 15.9 & 15.2 & 10.3 & 18.6 & 17.2 & 14.2 & 3.3 & 4.2 & 3.6 & 3.5\\
5000  & 11.0 & 12.2 &  8.5 & 14.4 & 12.3 & 10.2 & 2.2 & 2.9 & 2.5 & 2.1\\
10000 & 5.7  & 9.3  & 6.2  & 9.6  & 7.1  & 6.3 & 1.5  & 2.1 & 1.9 & 1.5\\
50000 & 1.3  & 1.3  & 1.2  & 2.6  & 2.4  & 2.0 & 0.7  & 0.8 & 0.7 & 0.7\\
\hline
\end{tabular}
\vspace{1ex}
\begin{tablenotes}
\item {\footnotesize Notes: Number of replications=1000. Results are rounded to 3 digits. Only estimated elements of $Pyd$ are displayed. Only linearly independent estimated elements of $\hat{F}^{\rumo}_1$ are displayed.}
\end{tablenotes}
\end{threeparttable}
\end{table}

\section{Estimation Results Omitted From the Main Text}\label{app: empirical application}
Given that the estimators of $\beta_D$ in Section~\ref{sec: empirical application} are GMM estimators, if we treat the estimated shares $\rands{s}_{Py,D,j,t}$ as the true shares (i.e. no estimation error), then we can easily construct the 2-step efficient GMM standard errors. Table~\ref{table: estimates of beta, appendix} displays the estimates of $\beta_D$ together with their standard errors.
\begin{table}[h]
\centering
\begin{threeparttable}
\centering
\caption{Estimates of $\beta$}\label{table: estimates of beta, appendix}
\begin{tabular}{lccc}
\hline
\hline
   &Naive & \{CTL, GM, K, O, Q\} & \{CTL, GM, K, O\}\\ 
$\hat{\beta}$ & -16.11 & -5.84 & -56.46\\
std. error & 2.30 & 7.1 & 20.04\\
\hline
\end{tabular}
\vspace{1ex}
\begin{tablenotes}
\item {\footnotesize Notes:Standard errors are computed assuming that there is no estimation error in shares. Results are rounded to 2 digits.}
\end{tablenotes}
\end{threeparttable}
\end{table}
Following \citet{nevo2001measuring}, here we report the median across markets own-price elasticities in Table~\ref{table: elast app}. In the first column we use estimates of the price coefficient assuming that there is no choice set variation (Naive). The second and third columns were constructed by using the price coefficients for those consumers who considered all 5 brands or $\{CTL, GM, K, O\}$, respectively. 

\begin{table}[h]
\centering
\begin{threeparttable}
\centering
\caption{Estimates of Median Across Markets Own-Price Elasticities}\label{table: elast app}
\begin{tabular}{lccc}
\hline
\hline
   &Naive & \{CTL, GM, K, O, Q\} & \{CTL, GM, K, O\}\\ 
CTL & -1.83 & -0.72 & -7.32\\
GM  & -2.41 & -0.86 & -9.51\\
K   & -2.04 & -0.7  & -8.15\\
O   & -2.07 & -0.77 & -7.83\\
Q   & -2.38 & -0.86 & 0\\
\hline
\end{tabular}
\vspace{1ex}
\begin{tablenotes}
\item {\footnotesize Notes: The first column is computed assuming that consumers face all 5 brands. The second column is computed assuming choice set variation for those consumers who consider all 5 brands. The last column is computed for those consumers who do not consider Quaker. Results are rounded to 2 digits.}
\end{tablenotes}
\end{threeparttable}
\end{table}
\end{document}